\immediate\write18{makeindex \jobname.nlo -s nomencl.ist -o \jobname.nls}

\documentclass[journal]{IEEEtran}
\usepackage{cite}
\usepackage{hyperref}

\usepackage{multirow}

\ifCLASSINFOpdf
   \usepackage[pdftex]{graphicx}
   \graphicspath{{../pdf/}{../jpeg/}}
   \DeclareGraphicsExtensions{.pdf,.jpeg,.png}
\else
   \usepackage[dvips]{graphicx}
   \graphicspath{{../eps/}}
   \DeclareGraphicsExtensions{.eps}
\fi
\usepackage{amsmath}
\usepackage{array}

\usepackage{enumerate,amssymb,epsfig,euscript,indentfirst,authblk}
\usepackage{subcaption}
\usepackage{booktabs}
\usepackage{nccmath}
\usepackage{lipsum}
\usepackage{mathtools, cuted}
\usepackage{graphics}
\usepackage{algorithm}
\usepackage{algpseudocode}
\usepackage{nomencl}
\makenomenclature

\usepackage{etoolbox}

\usepackage{soul,color}
\usepackage{gensymb}
\usepackage{svg}

\usepackage{makecell}

\begin{document}
%
\title{Hybrid Offline-online Scheduling Method for Large Language Model Inference Optimization}
%
%
%

\author{Bowen~Pang,
        Kai~Li,
        Ruifeng~She,
        and~Feifan~Wang,~\IEEEmembership{Member,~IEEE}
\thanks{B. Pang, K. Li, and R. She are with Noah's Ark Lab, Huawei.}
\thanks{F. Wang is with the Department
of Industrial Engineering, Tsinghua University, Beijing,
100084, China. E-mail: wangfeifan@tsinghua.edu.cn.}
}

\maketitle

\begin{abstract}
With the development of large language models (LLMs), it has become increasingly important to optimize hardware usage and improve throughput. In this paper, we study the inference optimization of the serving system that deploys LLMs. To optimize system throughput and maximize hardware utilization, we formulate the inference optimization problem as a mixed-integer programming (MIP) model and propose a hybrid offline-online method as solution. The offline method improves large-scale inference systems by introducing a Minimizing Makespan Bin Packing Problem. We further provide a theoretical lower bound computation method. Then, we propose an online sorting and preemptive scheduling method to better utilize hardware. In the online iteration scheduling process, a Lagrangian method is applied to evaluate the cost efficiency of inserting prefill stages versus decode stages at each iteration and dynamically determine when to preempt decoding tasks and insert prefill tasks. Experiments using real-world data from the LLaMA-65B model and the GSM8K dataset demonstrate that system utilization improves from 80.2\% to 89.1\%, and the total inference time decreases from 201.00 to 190.58 seconds. A 100-cases study shows that our method consistently outperforms the baseline method and improves the utilization rate by 8.0\% on average. Finally, we discuss potential future extensions, including stochastic modeling, reinforcement learning-based schedulers, and dynamic decision-making strategies for system throughput and hardware utilization.

\end{abstract}

\def\abstractname{Note to Practitioners}
\begin{abstract}
This work provides optimization tools for enhancing the efficiency of LLM inference systems through advanced scheduling techniques. From the perspective of LLM inference service providers, improved hardware utilization can reduce operational costs by requiring less hardware to maintain the same level of service. From the user’s perspective, reduced inference time translates to faster response times and improved service quality. Furthermore, the proposed scheduling techniques are adaptable to various LLM models, hardware platforms, and datasets, making them highly scalable and broadly applicable to real-world LLM inference scenarios.
\end{abstract}


%
\IEEEpeerreviewmaketitle

\section{Introduction}
\label{introduction}
Recent advancements in large language models (LLMs), including GPT-4, LLaMA, and Qwen, have significantly transformed the landscape of natural language processing by enabling more sophisticated text generation, comprehension, and interaction capabilities. These models serve as foundational technologies in a wide range of applications, such as chatbots, machine translation, and content creation. Despite their transformative potential, the inference process for LLMs is computationally intensive and resource-demanding, resulting in high costs and increased latency. These factors pose significant challenges when scaling their deployment across various applications. Effective inference scheduling is crucial for optimizing resource usage, reducing operational costs, and ensuring high-quality service delivery.

\begin{figure}
    \centering
    \includegraphics[width=0.98\linewidth]{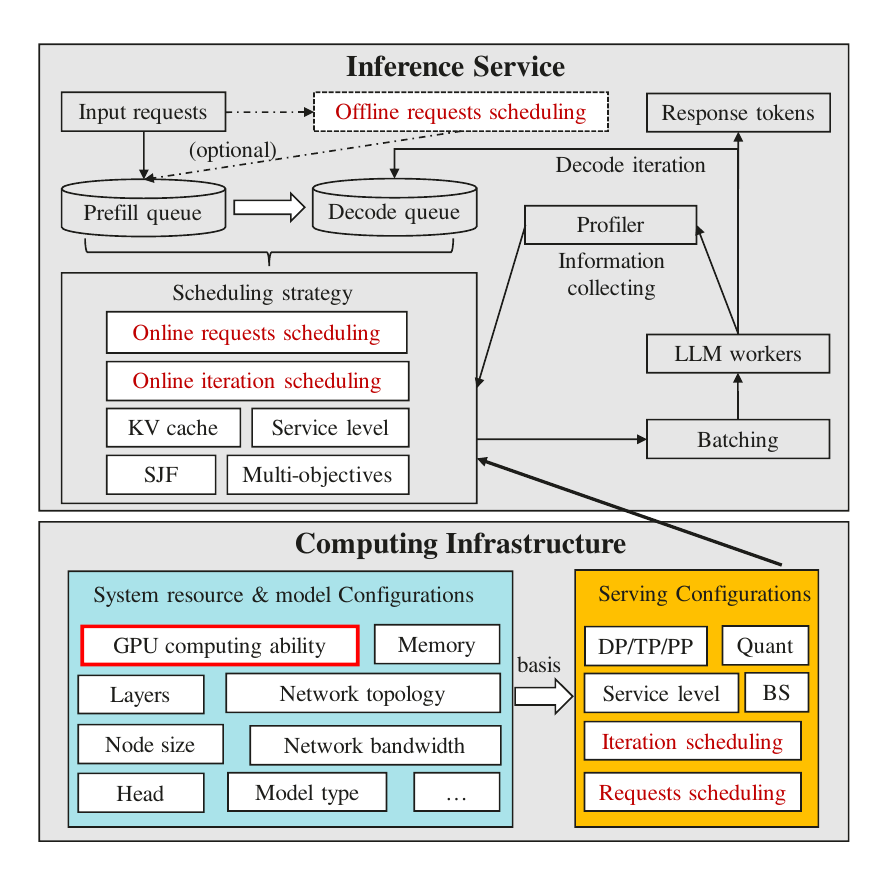}
    \caption{Illustration of LLM inference service system}
    \label{fig:intro_inference_service}
\end{figure}

As illustrated in Fig. \ref{fig:intro_inference_service}, the current inference service system comprises three components: system resource, serving configuration, and inference service. System resource determines the extent of computational resources available in the system including GPU computing ability, memory, node size, network bandwidth, etc. The serving configuration specifies the desired model deployment method and service level expectations, such as data/tensor/pipeline parallel settings, quantization, batch size, and scheduling configurations. The inference service processes input requests from users, utilizes hardware capabilities, fulfills user configurations, and returns responses to LLM users. The scheduling strategy module is the core of the inference service, managing prefill and decode queues, offline and online request scheduling, and iteration-level hardware management. The offline scheduling method is optional, only for inference tasks where all requests are known in advance. Conversely, the two online scheduling methods, i.e., requests scheduling and iteration scheduling, are more versatile and applicable to a wide range of scenarios. The online methods acquire information from the online profiler and dispatch requests and inference tasks to LLM workers. In practice hardware utilization includes up to 20\% ``bubbles'', which means the hardware is idle during inference service, and detail is later shown in Section \ref{solution_method}. By designing an efficient LLM inference scheduling system to reduce these bubbles, computational resource consumption can be decreased, leading to reduced latency and increased throughput. For maintaining logic flow, detailed technical introductions to prefill and decode operations are provided in Section \ref{problem_formulation}.

Current works, such as vLLM \cite{kwon2023efficient} and ORCA \cite{yu2022orca}, provide a robust foundation for LLM inference serving systems. The primary contributions of vLLM and ORCA to LLM inference are their enhancements in resource allocation and execution efficiency, which significantly increase throughput and decrease latency in large-scale language model deployments. These improvements are achieved through advanced memory management and continuous batching techniques, which enhance model parallelism and effectively leverage hardware resources \cite{kwon2023efficient, yu2022orca}.
However, the state-of-the-art scheduler in vLLM predominantly employs a First-Come-First-Serve (FCFS) and prefill-first policy, prioritizing prefill tasks but not fully leveraging the scheduling system's potential. We identify the causes of low hardware utilization rates. From an offline scheduling perspective, client loads are unbalanced, and request sequencing can be improved. From an online scheduling angle, the prefill-first policy lacks parallel processing of multiple requests at prefill stage, and there are no dynamic sequencing adjustments or preemption methods.

The optimization of LLM inference presents two major challenges. The first challenge pertains to the offline method, which involves managing a large number of requests from up to 200 parallel clients. This task is particularly time-consuming when using a mixed-integer programming (MIP) scheduling model. The second challenge is the need for much faster decision-making for online methods compared to traditional online scheduling problems. For example, in LLM inference scenarios, online decisions about whether to send prefill or decode requests to LLM workers typically occur every 50 milliseconds. In contrast, in healthcare or production systems, online decisions usually occur every thirty seconds or even several minutes. For online scheduling methods, the higher frequency of decision-making requires algorithms that are efficient and capable of delivering results within extremely short time frames.


Accompanied by these challenges, developing a scheduling method for LLM inference could yield substantial benefits. According to NVIDIA Financial Results in 2024 \cite{nvidia2024finance}, the revenue from GPU retail and cloud computation service has reached \$60.9 billion per year. Improving computational resource efficiency by 5\% will earn up to \$3.05 billion in revenue annually. Therefore, research in this area is crucial, especially for scheduling researchers.

In this paper, we aim at enhance service system throughput and optimize hardware utilization. We define the LLM inference optimization problem using a MIP model as its uncertainty equivalence. Then, we propose a hybrid offline-online method as solution. This work is the first to define this problem from a scheduling perspective. To address the high real-time demands in LLM inference, we propose a hybrid offline-online scheduling method. In numerical experiments, we demonstrate that our method can improve system throughput by 5.46\% and improve hardware utilization by 11.0\%. A 100-cases study shows that our method outperforms baseline method consistently with an improvement of 8.0\% on utilization rate.

The major contributions of this work include the followings. To the best of our knowledge, we are the first to formulate a mathematical model that describe the scheduling problem in the area of LLM inference. We design a two-stage approach to manage both offline and online scheduling challenges. An Minimizing Makespan Bin Packing model is developed to efficiently solve the offline scheduling problem. Additionally, we introduce a sorting and preemption method to handle the online request scheduling problem, and we develop a Lagrangian-based heuristic technique for solving the online iteration scheduling issue. In the online scheduling module, our method can provide decisions within 5 milliseconds, meeting the real-time decision-making requirements of practical application in LLM inference.

The remainder of the paper is structured as follows. We review literature on efficient LLM inference in Section \ref{literature}. The problem definition and model formulation are presented in Section \ref{problem_formulation}. Then, we illustrate the difficulty of solving this problem and introduce our hybrid offline-online scheduling method to provide a timely solution in Section \ref{solution_method}. Numerical studies using real cases and 100 generated cases are presented in Section \ref{numerical-experiment}. Finally, the conclusion and future directions are provided in Section \ref{conclusion}.
\section{Literature review}
\label{literature}


\begin{table*}[t]
    \centering
    \caption{Literature Review}
    \begin{tabular}{cccccccccc}
    \toprule
        Work & Throughput & Latency & \makecell{Cache \\ Management} & \makecell{Dynamic \\ Decision} & Uncertainty & \makecell{Request \\ Management} & \makecell{Iteration\\ Management} & Batching & \makecell{Distributed \\ Strategies}\\
    \midrule
        Orca & \checkmark & \checkmark &  &  & \checkmark  &  & \checkmark & \checkmark & \\
        vLLM & \checkmark  & \checkmark  & \checkmark  &  &  &  &  &  & \\
        Sarathi-Serve\cite{agrawal2024sarathi} & \checkmark & \checkmark &  &  &  &  & \checkmark & \checkmark & \\
        Llumnix\cite{sun2024llumnix} &  & \checkmark & \checkmark & \checkmark &  & \checkmark &  &  & \checkmark\\
        InfiniGen\cite{lee2024infinigen} &  & \checkmark & \checkmark &  &  &  &  &  & \\
        dLoRA\cite{wu2024dlora} & \checkmark & \checkmark  & \checkmark & \checkmark &  &  &  & \checkmark & \checkmark\\
        VTC\cite{sheng2024fairness} &  &  &  & \checkmark &  & \checkmark &  &  & \\
        FastServe\cite{wu2023fast} & \checkmark  & \checkmark & \checkmark & \checkmark & \checkmark & \checkmark &  &  & \\
        DistServe\cite{zhong2024distserve} & \checkmark & \checkmark & \checkmark & \checkmark &  &  & \checkmark &  & \checkmark\\
        MoonCake & \checkmark & \checkmark & \checkmark & \checkmark &  &  &  &  & \checkmark \\
    \midrule
        OURS & \checkmark & \checkmark & \checkmark & \checkmark & \checkmark & \checkmark & \checkmark & \checkmark & \\
    \bottomrule
    \end{tabular}

    \label{table:literature_review}
\end{table*}

This section provides an overview of existing research and prevalent methodologies in inference optimization for LLMs. Firstly, we introduce the general techniques commonly employed in model serving, which can be seamlessly integrated with our scheduling strategies. Subsequently, we elucidate several classical techniques widely adopted in LLM inference systems, which constitute the cornerstone of our framework and methodologies. In the following, we succinctly introduce recent advancements in inference optimization. Finally, we examine scheduling methods within the existing operations research domain.

As LLM inference falls within the broader scope of model serving, a variety of general inference optimization techniques can be effectively utilized. Model compression is one of the quintessential optimization strategies for reducing model size, encompassing techniques such as quantization \cite{jacob2018quantization}, sparsification \cite{child2019generating,bai2024sparsellm}, and distillation \cite{hinton2015distilling}. In addition, the design of more compact structures to replace the original ones is also common. For instance, employing multi-query attention \cite{shazeer2019fast} or grouped-query attention \cite{ainslie2023gqa} in place of the original multi-head attention in Transformer architecture can reduce key-value heads, resulting in a more streamlined model. Nevertheless, both model compression and the design of compact structures can alter model weights, potentially leading to a decline in accuracy. Instead of optimizing the model size, data parallelism (DP) and model parallelism aim to fully leverage the computational power of the devices. In DP \cite{narayanan2021efficient}, the model weights are replicated across multiple devices, allowing different inference requests to be processed in parallel on different devices. Model parallelism distributes the model weights across several devices to minimize the per-device memory footprint of the model weights. Consequently, each device can operate more efficiently by executing a smaller portion of the task. Several model parallelization methods exist, such as pipeline parallelism (PP) \cite{huang2019gpipe}, tensor parallelism (TP) \cite{shoeybi2019megatron}, sequence parallelism (SP) \cite{korthikanti2023reducing}, and context parallelism (CP) \cite{liu2023ring}. Since our scheduling methods are orthogonal to the aforementioned techniques, both model optimization and parallelization strategies can be employed seamlessly in conjunction with our methods to enhance inference efficiency.



In addition to general model serving techniques, the optimization of LLM inference serving systems primarily involves enhancing the model forward pass. Researchers have improved system efficiency from various perspectives, including kernel fusion, Key-Value (KV) cache management, request management, iteration management, batching, distributed strategies, etc. Here, we present several classic techniques that are prevalent in LLM inference serving systems. FlashAttention \cite{dao2022flashattention} amalgamates the operations of data transfer between hardware components within the attention mechanism to expedite operation execution without compromising model accuracy. Speculative decoding \cite{leviathan2023fast, chen2023accelerating} employs an auxiliary model to generate a preliminary draft, followed by a verification process executed by the main model. This technique enables the serving system to output multiple tokens in a single forward pass instead of one. Orca \cite{yu2022orca} pioneers continuous batching by aggregating different requests at the iteration level. Rather than awaiting the completion of an entire batch before starting the execution of new requests, continuous batching allows new requests to be inserted into the batch while other requests are still in progress. Inspired by memory management strategies in operating systems, vLLM \cite{kwon2023efficient} introduces PagedAttention, wherein the attention key and value vectors are stored as non-contiguous blocks in memory. Continuous batching and PagedAttention significantly increase overall GPU memory utilization during the execution of LLM. SarathiServe \cite{agrawal2024sarathi} introduces chunked-prefills, also known as dynamic SplitFuse or prefill-decode (PD) Fusion, batching together prefill and decode chunks to maximize both computation and bandwidth utilization. Since serving systems are commonly deployed on distributed platforms, numerous strategies have been proposed to exploit distributed characteristics. For example, recent works \cite{patel2024splitwise,zhong2024distserve,hu2024inference} advocate for separating prefill servers from decode servers, also known as PD Separation, due to the distinct computational and bandwidth characteristics of these two stages. For a comprehensive review of these techniques, we recommend \cite{zhou2024survey} for further reference.
These classic techniques form the foundation of inference services and our methods. 

Recently, with ongoing advancements in AI system research, numerous innovative inference techniques have been developed, particularly those related to schedulers. We present some representative works and highlight the differences between these approaches and our method in TABLE~\ref{table:literature_review}. Inspired by context switching across CPU cores, Llumnix \cite{sun2024llumnix} proposes a live migration mechanism and a dynamic scheduling policy to reschedule requests across multiple model instances of LLM deployed on GPU clusters, thereby enhancing load balancing and isolation. InfiniGen \cite{lee2024infinigen} addresses the challenge of large KV cache sizes for long-text generation by speculating and prefetching critical KV cache entries. For LoRA models, dLoRA \cite{wu2024dlora} addresses the challenges of serving multiple LoRA models by dynamically merging and unmerging adapters with the base model and migrating requests and adapters between replicas. Sheng et al. \cite{sheng2024fairness} study the fairness problem in LLM serving concerning clients and proposes a novel scheduling algorithm called the virtual token counter (VTC). FastServe \cite{wu2023fast} proposed an innovative skip-join MLFQ scheduler to enable preemption during the autoregression process of LLM inference. In distributed systems, DistServe \cite{zhong2024distserve} tackles the issues of PD interference and resource coupling by disaggregating prefill and decoding computation, and proposes placement algorithms to optimize resource allocation and parallelism strategies for different phases. Mooncake \cite{qin2024mooncake} also proposes a disaggregated architecture that separates the prefill and decode clusters and utilizes a KV cache-centric scheduler to manage the cache flow. While these innovative techniques attempt to address the issues faced by the scheduler to some extent, they scarcely model the scheduling problem formally and are limited to solve the inference optimization problem theoretically.

In the domain of operations research, scheduling is a well-established and extensively utilized approach. For instance, offline scheduling methods are applied in manufacturing systems \cite{wang2020print3D}, healthcare systems \cite{pang2018surgery}, and operations management systems \cite{erdogan2015online}. To address real-time decision-making or multi-stage stochastic scenarios, online scheduling methods have been introduced in these systems \cite{pang2022dynamic, lee2019online}. Nevertheless, none of the systems we examined achieve the rapid decision frequency—down to 10 milliseconds—that is observed in LLM inference systems. Furthermore, traditional stochastic programming models are not suitable for sequential decision-making problems where real-time information is continuously revealed. Therefore, it is imperative to develop and tailor traditional methods for application in this emergent domain of LLM inference. Research on LLM inference optimization can not only enhance hardware resource utilization in the LLM domain but also expand the repertoire available to existing operations research algorithms.


\section{Problem formulation}
\label{problem_formulation}

\subsection{Background of LLM inference techniques}
High efficiency and low latency are critical in LLM inference, and they depend not only on performance of hardware, such as GPU, but also on how well the hardware is used. As pushing the limit of hardware computing power is costly and faced with slow progress, improved inference scheduling becomes a promising means of achieving the same outcome. Three factors are involved when designing inference scheduling methods: PD management approaches, cache management, and batching strategy.

\begin{figure}
    \centering
    \includegraphics[width=0.98\linewidth]{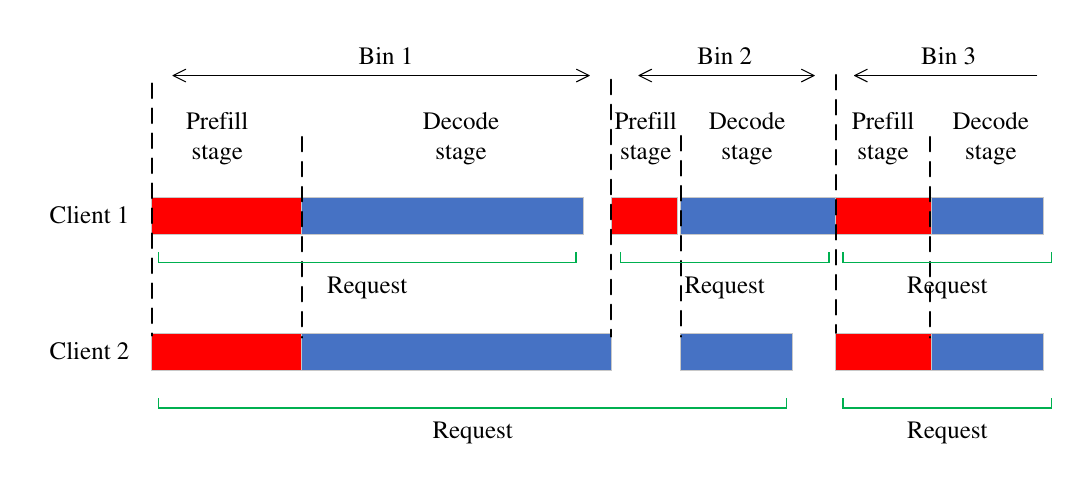}
    \caption{Illustration of LLM inference}
    \label{fig:illustration}
\end{figure}

LLM inference alternately goes through two stages, i.e., prefill and decode stages. In the prefill stage, initial input tokens are processed, and the LLM model's internal states, such as hidden states, are prepared. In the decode stage, the LLM model generates output tokens based on the context established during the prefill stage. The time lengths for prefill stage and decode stage are uncertain. The alternating process continues in parallel, until either the end of the sequence is reached or a predefined stopping criterion is satisfied. The process of LLM inference is illustrated in Fig. \ref{fig:illustration}, where requests, each with a prefill phase and a decode phase, are sent to clients and processed in parallel. The whole process is divided into multiple bins, and each bin consists of a prefill stage and a decode stage. Requests in prefill phase can be served only when the process is in prefill stage, and the same requirement is applied to the decode phase and stage. A client may be idle as either the prefill phase or decode phase is completed but the process still stays in the same stage, causing compromised utilization of computing resources. A single LLM inference in practice typically contains a large number of requests processed by parallel clients. Thus, there is great potential in better utilizing computing resources in LLM inference, but the scheduling problem has a high complexity and, as a real-time decision making in milliseconds, the computing time is limited. The most commonly used approaches to managing prefill are provided below.

\begin{figure}
    \centering
    \includegraphics[width=0.98\linewidth]{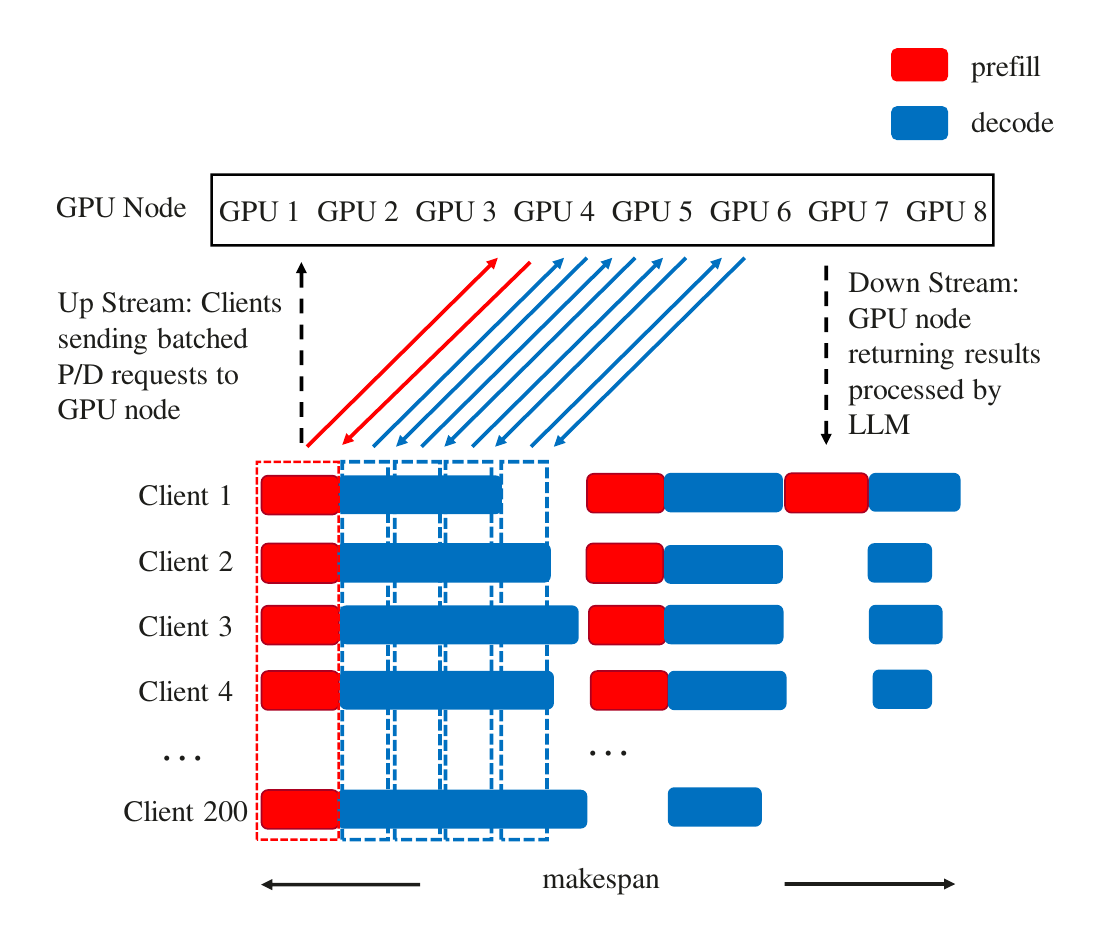}
    \caption{Illustration of PD Competition}
    \label{fig:PDcompetition}
\end{figure}

\begin{itemize}
    \item PD Competition: As illustrated in Fig. \ref{fig:PDcompetition}, in this approach, either prefill or decode stage is processed at any given time for all clients on a single hardware node (typically consisting of 8 GPUs). PD Competition allows decode stage to be preempted by the prefill stage to enhance hardware utilization.
    \item PD Fusion: This approach integrates the prefill and decode stages into a single cohesive operation, aimed at reducing overhead and enhancing throughput by streamlining the inference pipeline. This approach also attempts to decrease latency through alignment of processes. However, this integration compromises flexibility, restricting the ability to independently optimize each process or tailor responses to varying workload demands.
    \item PD Separation: This approach separates the prefill and decode stages across exclusive sets of GPUs. However, it introduces additional communication or coordination overhead, which increases latency if not properly managed.

\end{itemize}

As a widely used approach, PD Competition has a high flexibility in effectively utilizing computing resources. Such an approach also allows an inference scheduling method to fit in and further enhance its performance. As is aforementioned, inference scheduling for LLM inference is challenging. This study focuses on the inference scheduling under the PD Competition approach.

The second factor that influences the efficiency of LLM inference is cache management. The KV cache is instrumental during the decoding stage by storing intermediate hidden states from preceding token generation steps. It allows the LLM model to reuse states, significantly accelerating the inference process. Despite its advantages, the KV cache requires to be properly managed. First, the KV Cache size increases with the length of input and output sequences and the number of LLM model layers. This cache size growth results in significant memory consumption, especially in the context of large-sized models and extended sequences. Effective KV cache management avoids memory overflow and sustains high inference speed. Cache management may involve caching only the most relevant hidden states, while discarding or compressing less critical information to optimize resource use. Second, concurrency issues should be addressed. The complexity of managing the KV cache escalates with concurrent inference tasks. Ensuring consistent and conflict-free partitioning and access to the cache for each task is important for performance and accuracy of LLM inference. Besides, although the KV cache alleviates computational load during decoding, it introduces cache access and management overheads. Cache management requires taking into account overall latency. In this study, we proactively compute the KV cache and determine the optimal maximum number of parallel requests, equivalent to the number of clients handled in subsequent stages. Thus, we assume in the inference scheduling problem shown in Fig. \ref{fig:illustration} that the total number of clients is predetermined.

The third factor that influences LLM inference efficiency is batching strategy. Continuous batching, which is introduced by ORCA \cite{yu2022orca}, has emerged as an innovative technique to enhance the inference efficiency by dynamically aggregating incoming requests in real time. Unlike traditional static batching, which waits for a fixed number of requests before processing, continuous batching minimizes latency by reducing idle times between batch executions and adapting to fluctuating workloads. It ensures efficient use of computational resources. By maximizing parallel processing capabilities and aligning with the model architecture's latency and throughput trade-offs, continuous batching significantly enhances the scalability and responsiveness of LLM deployments. Using the continuous batching technique, decoding is allowed to be preempted by the prefill stage. Such a case is shown by the first request of client 2 in Fig. \ref{fig:illustration}, where the decode phase is separated by the prefill stage of bin 2 and is assigned to both the decode stages of bin 1 and 2. In this work, the decision-making process allows decoding to be preempted by the prefill stage, facilitated by the continuous batching technique.

\begin{table*}
    \centering
    \caption{Notation Table: Sets, Parameters and Decision Variables}
    \begin{tabular}{cl}
    \toprule
        Sets    & Description \\
    \midrule
        $\mathcal{I}=\left\{1,2,\cdots \right\}$  & Set of requests. \\
        $\mathcal{J}=\left\{1,2,\cdots \right\}$  & Set of clients.\\
        $\mathcal{K},\mathcal{K}^\text{p}, \mathcal{K}^\text{d}=\left\{1,2,\cdots \right\}$& Set of bins, prefill stages, and decode stages, respectively. $\mathcal{K} = \mathcal{K}^\text{p} = \mathcal{K}^\text{d}$.\\
        $\mathcal{L}=\left\{1,2,\cdots \right\}$  & Set of all possible levels for the prefill stage.\\
    \midrule
        Parameters          & \\
    \midrule
        $I=\mid \mathcal{I} \mid $                 & Total number of requests.\\
        $J=\mid \mathcal{J} \mid $                 & Total number of clients.\\
        $K=\mid \mathcal{K} \mid $                 & Total number of bins.\\
        $N^\text{p}_i \in \mathbb{Z}^+$             & Input token number of request $i$, for $i\in \mathcal{I}$. \\
        $N^\text{d}_i \in \mathbb{Z}^+$         & Output token number of request $i$, for $i\in \mathcal{I}$. \\
        $N^\text{cap}_l \in \mathbb{Z}^+$        & Maximum token capacity of level $l$, for $l\in \mathcal{L}$. \\  
        $T^\text{d} \in \mathbb{R}^+$            & Decode time per token.\\
        $T^\text{p} \in \mathbb{R}^+$            & Prefill time per token. \\
        $T^\text{p}_l \in \mathbb{R}^+$          & Prefill time of level $l$, for $l\in \mathcal{L}$. \\
    \midrule
        Decision Variables      & \\
    \midrule
        $p_{i,j,k}\in \{0,1\}$ & Assignment of prefill stage $k$ to request $i$ on client $j$ for prefill phase, for $i\in \mathcal{I}$, $j\in \mathcal{J}$, and $k\in \mathcal{K}^\text{p}$.  \\
        $d_{i,j,k}\in \{0,1\}$ & Assignment of decode stage $k$ to request $i$ on client $j$ for decode phase, for $i\in \mathcal{I}$, $j\in \mathcal{J}$, and $k\in \mathcal{K}^\text{d}$.  \\
        $t^\text{s,p}_{k}\in \mathbb{R}^+$  & Start time of the $k$th prefill stage, for $k\in \mathcal{K}^\text{p}$.  \\
        $t^\text{s,d}_{k}\in \mathbb{R}^+$  & Start time of the $k$th decode stage, for $k\in \mathcal{K}^\text{d}$.\\
        $n^\text{p}_{k} \in \mathbb{R}^+$ & Time length of the $k$th  prefill stage, for $k\in \mathcal{K}^\text{p}$.  \\
        $n^\text{d}_{k} \in \mathbb{R}^+$ & Time length of the $k$th  decode stage, for $k\in \mathcal{K}^\text{d}$.  \\
        $w_{i,j,k}\in [0,1]$   & The proportion of the decoding phase of request $i$ executed in the $k$th decode stage on client $j$, for $i\in \mathcal{I}$, $j\in \mathcal{J}$, and $k\in \mathcal{K}^\text{d}$. \\
        $x_{i,j}\in \{0,1\}$ & The assignment of request $i$ to client $j$, for $i\in \mathcal{I}$ and $j\in \mathcal{J}$. \\
        $y_{k,l}\in \{0,1\}$    & Indicator variable specifying if the $k$th prefill stage is at level $l$, for $k\in \mathcal{K}^\text{p}$ and $l\in \mathcal{L}$.\\
        $t^\text{max} \in \mathbb{R}^+$ & Total inference time for all requests completed.\\
    \bottomrule
    \end{tabular}
    \label{notation table}
\end{table*}

\subsection{Inference scheduling problem description}
Inference scheduling aims to schedule inference requests using continuous batching and PD Competition, with the goal of minimizing the total inference time while adhering to operational constraints. The problem settings are given as follows and related notations are presented in TABLE \ref{notation table}.

\begin{figure}
    \centering
    \includegraphics[width=0.9\linewidth]{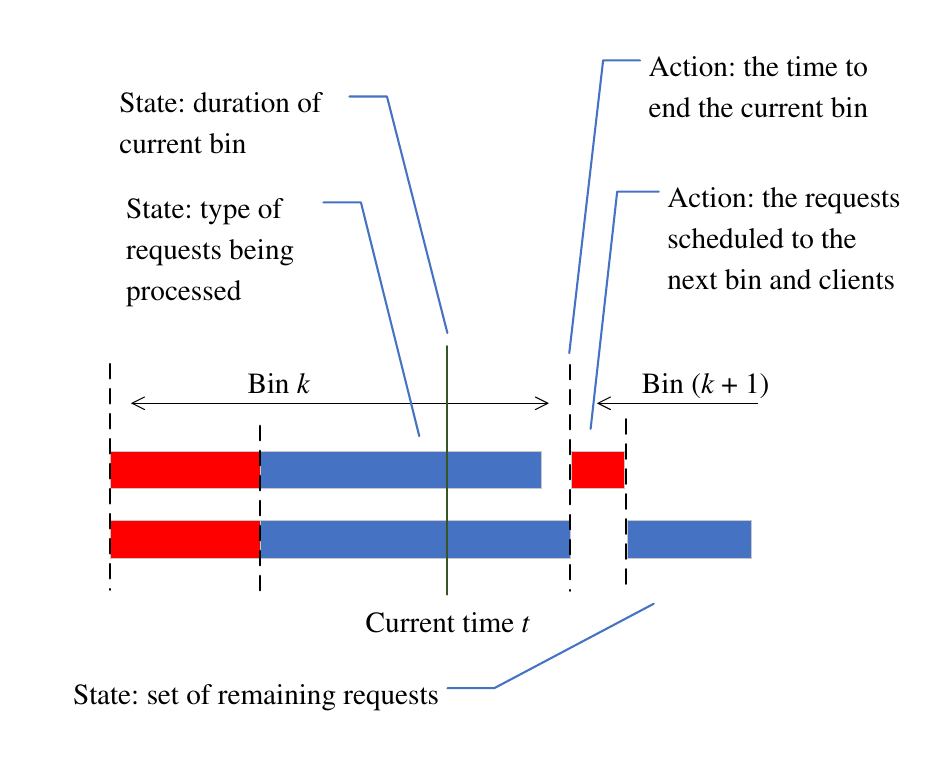}
    \caption{Online scheduling}
    \label{fig:control}
\end{figure}

\begin{enumerate}
    \item \textbf
{Requests and processing time:}
    \begin{itemize}
        \item Let \( \mathcal{I}=\left\{ 1,2,\cdots\right\} \) be the set of inference requests. Inference request $i$, for $i \in \mathcal{I}$, has a fixed input token number $N^\text{p}_i \in  \mathbb{Z}^+$ for prefill phase and output token number $N^\text{d}_i \in  \mathbb{Z}^+$ for decode phase. The input token number is assumed to be known, but the output token number is unknown.
        \item Each request has a known prefilling time, linearly related to the total number of input tokens in a bin.
        \item The decoding time is approximately linearly related to the output token number. Let $\mathcal{J}=\left\{1,2,\cdots \right\}$ be the set of clients and $T^\text{d}$ be the decode time per token. The minimal unit of decoding time is equal to the amount of time that each client processes a single token, i.e., $T^\text{d}|\mathcal{J}|$.
    \end{itemize}
    \item \textbf{Bin and batch size:}
    \begin{itemize}
        \item Let $\mathcal{K}=\left\{1,2,\cdots \right\}$ be the set of bins. The inference process is divided into a total of \( K=|\mathcal{K}| \) bins.
        \item The batch size \( |\mathcal{J}| \) represents the maximum number of requests that can be processed simultaneously in a batch.
    \end{itemize}
    \item \textbf{Stages:}
    \begin{itemize}
        \item There are two stages in hardware operation system: prefill and decode. Prefill and decode stages must alternate, ensuring that each stage is dedicated exclusively to one type of operation.
        \item At any given time, the system can perform either prefill or decode stages, but not both.
        \item At any time as the system starts a decode stage, the length of this operation is determined. Meanwhile, all requests processed at this stage will be scheduled. The decision is made based on system state, including information of the duration of the current bin, type of requests being processed, and the set of remaining requests. It is illustrated in Fig. \ref{fig:control}.
    \end{itemize}
    \item \textbf{Assignment and allocation:}
        \begin{itemize}
        \item Each request must be assigned to exactly one prefill stage for processing.
        \item For every request, the prefill phase must be completed by the same client that will subsequently carry out its decode phase. Hence, both phases must be allocated to the same client.
        \item  A client can process only one request at a time. Once a client begins processing a request, it remains occupied and cannot preempt tasks until both the prefill and decode stages of that request are completed.
    \end{itemize}
\end{enumerate}

\subsection{Deterministic equivalence}
In this hybrid offline-online problem, the offline decision component involves determining the assignment and sequence of given requests on clients.
As a sequential decision-making problem, the online decision component focuses on determining the length of each bin and the sequence of remaining requests in the future, with the aim of minimizing the total inference time. 

Without considering uncertainty, the offline-online inference scheduling can be formulated as a deterministic equivalence, which is a form of an MIP model as follows. 

\begin{align}
    \label{obj:makespan}
    & \min t^\text{max} \\
    & \nonumber \text{s.t.} \\
    \label{cst:makespan_def}
    & t^\text{max} \geq t^\text{s,d}_{k} + n^\text{d}_{k}, \text{ for } k \in \mathcal{K}^\text{d}, \\
    \label{cst:bin_def_2}
    & t^\text{s,p}_{k} - \left(t^\text{s,d}_{k-1} + n^\text{d}_{k-1}\right) \geq 0, \text{ for } k = 2,...,K,\\
    \label{cst:bin_def_3}
    & t^\text{s,d}_{k} - \left(t^\text{s,p}_{k} + n^\text{p}_{k}\right) \geq 0, \text{ for } k = 1,2,...,K,\\
    \label{cst:prefill_length_def_1}
    & n^\text{p}_{k} \geq \sum_{l \in \mathcal{L}} T^\text{p}_l y_{k,l}, \text{ for } k \in \mathcal{K}^\text{p},\\
    \label{cst:prefill_length_def_2}
    & \sum_{i \in \mathcal{I}} \sum_{j \in \mathcal{J}} N^\text{p}_i p_{i,j,k} \leq \sum_{l \in \mathcal{L}} N^\text{cap}_l y_{k,l}, \text{ for } k \in \mathcal{K}^\text{p},\\ 
    \label{cst:level_def}
    & \sum_{l \in \mathcal{L}} y_{k,l} = 1,  \text{ for } k \in \mathcal{K}^\text{p}, \\
    \label{cst:decode_length_def_1}
    & n^\text{d}_{k} \geq T^\text{d} \sum_{i\in \mathcal{I}} N_i^\text{d} w_{i,j,k}, \text{ for }j \in \mathcal{J}, \text{ and } k \in \mathcal{K}^\text{d},\\
    \label{cst:immediate_prefill_decode}
    & d_{i,j,k} - p_{i,j,k} \geq 0, \text{ for }i \in \mathcal{I}, j \in \mathcal{J}, \text{ and } k \in \mathcal{K}^\text{d}, \\
    \label{cst:consecutive_decode_bin}
    \begin{split}
    &M\left(2 - d_{i,j,k_1} - d_{i,j,k_2}\right) + \sum_{k'=k1}^{k_2} d_{i,j,k'} \\ 
    &\quad \geq k_2 - k_1 + 1, \\
    &\quad \text{ for }i \in \mathcal{I}, j \in \mathcal{J}, k_1,k_2 \in \mathcal{K}^\text{d},\text{ and }k_1<k_2, 
    \end{split} \\
    \label{cst:prefill-decode-order}
    \begin{split}
    & M\left(p_{i,j,k_1}-1\right) +d_{i,j,k_2} \leq 0, \\
    &\quad \text{ for }i \in \mathcal{I}, j \in \mathcal{J}, k_1 \in \mathcal{K}^\text{p},k_2\in \mathcal{K}^\text{d}, \text{ and }k_1 > k_2, 
    \end{split} \\
    \label{cst:d_def_1}
    & \sum_{i\in \mathcal{I}} d_{i,j,k} \leq 1, \text{ for }j \in \mathcal{J}, \text{ and } k \in \mathcal{K}^\text{d}, \\
    \label{cst:d_def_2}
    & \sum_{k \in \mathcal{K}^\text{d}} d_{i,j,k} \leq K, \text{ for }i \in \mathcal{I}, \text{ and }j \in \mathcal{J}, \\
    \label{cst:w_def_1}
    & \sum_{k \in \mathcal{K}^\text{d}} w_{i,j,k} = x_{i,j}, \text{ for }i \in \mathcal{I}, \text{ and }j \in \mathcal{J}, \\
    \label{cst:w_def_2}
    & \sum_{j\in \mathcal{J}} \sum_{k \in \mathcal{K}^\text{d}} w_{i,j,k} = 1, \text{ for }i \in \mathcal{I}, \\
    \label{cst:p_def_1}
    & \sum_{i \in \mathcal{I}} p_{i,j,k} \leq 1, \forall j, k, \\
    \label{cst:p_def_2}
    & \sum_k p_{i,j,k} = x_{i,j}, \text{ for }j \in \mathcal{J}, \text{ and } k \in \mathcal{K}^\text{p}, \\
    \label{cst:loc_def}
    & \sum_{j \in \mathcal{J}} x_{i,j} = 1, \text{ for }i \in \mathcal{I}, \\
    \label{cst:var_def_start}
    & x_{i,j} \in \{0,1\}, \text{ for }i \in \mathcal{I}, \text{ and } j \in \mathcal{J}, \\
    & y_{k,l} \in \{0,1\}, \text{ for }k \in \mathcal{K}^\text{p}, \text{ and } j \in \mathcal{J}, \\
    & p_{i,j,k} \in \{0,1\}, \text{ for }i \in \mathcal{I}, j \in \mathcal{J}, k \in \mathcal{K}^\text{p}, \\
    & d_{i,j,k} \in \{0,1\}, \text{ for }i \in \mathcal{I}, j \in \mathcal{J}, k \in \mathcal{K}^\text{d}, \\
    & w_{i,j,k} \in [0,1], \text{ for }i \in \mathcal{I}, j \in \mathcal{J}, k \in \mathcal{K}^\text{d}, \\
    & t^\text{s,p}_{k}, n^\text{p}_{k}, \text{ for }k \in \mathcal{K}^\text{p},\\
    \label{cst:var_def_end}
    & t^\text{s,d}_{k}, n^\text{d}_{k}, \text{ for }k \in \mathcal{K}^\text{d}.
\end{align}
The total inference time is denoted by $t^\text{max}$, which is minimized in the objective function given in Eq. (\ref{obj:makespan}). We let $\mathcal{K}^\text{d}=\left\{1,2,\cdots \right\}$ be the set of decode stages. For any decode stage $k\in \mathcal{K}^\text{d}$, let $t^\text{s,d}_{k}\in \mathbb{R}^+$ and $n^\text{d}_{k}\in \mathbb{R}^+$ denote the start time and time length of the $k$th decode stage, respectively, and thus its end time is $\left(t^\text{s,d}_{k}+n^\text{d}_{k}\right)$. Eq. (\ref{cst:makespan_def}) requires that the total inference time is greater than or equal to the end time of any decode stage. Let $K \in \mathbb{Z}^+$ be the total number of bins, prefill stages, and also decode stages. Any prefill stage should start after the end of its previous decode stage, suggested by Eq. (\ref{cst:bin_def_2}). For any prefill stage $k\in \mathcal{K}^\text{p}$, let $n^\text{p}_{k}\in \mathbb{R}^+$ be the time length of the $k$th prefill stage. Thus, Eq. (\ref{cst:bin_def_3}) suggests that, within the same bin, the decode stage should start after the end of prefill stage. We use $\mathcal{L}=\left\{1,2,\cdots \right\}$  to represent the set of all possible levels for a prefill stage, and the time length of a prefill stage depends on the level. Let $y_{k,l}\in \left\{0,1\right\}$, for $k\in \mathcal{K}^\text{p}$ and $l \in \mathcal{L}$, be an indicator variable. Prefill stage $k$ is at level $l$, if $y_{k,l}=1$. Otherwise, the prefill stage is not at level $l$. We denote by $T^\text{p}_l \in \mathbb{R}^+$ the prefill time of level $l$, for $l\in \mathcal{L}$, and thus the time length of a prefill stage is determined in Eq. (\ref{cst:prefill_length_def_1}). We use $N^\text{cap}_l \in \mathbb{Z}^+$ to denote the maximum token capacity of level $l$, for $l\in \mathcal{L}$, and let $p_{i,j,k}\in \left\{0,1\right\}$ denote assignment of prefill stage $k$ to request $i$ on client $j$ for prefill phase, for $i\in \mathcal{I}$, $j\in \mathcal{J}$, and $k\in \mathcal{K}^\text{p}$. The relationship between $p_{i,j,k}$ and $y_{k,l}$ is given in Eq. (\ref{cst:prefill_length_def_2}). Eq. (\ref{cst:level_def}) suggests that any prefill stage can only be at one level in $\mathcal{L}$. Different than the prefill phase, the decode phase of a request can be served in multiple decode stages. Thus, we introduce $w_{i,j,k}\in [0,1]$ to represent the proportion of the decoding phase of request $i$ executed in the $k$th decode stage on client $j$, for $i\in \mathcal{I}$, $j\in \mathcal{J}$, and $k\in \mathcal{K}^\text{d}$. The time length of a decode stage is provided by Eq. (\ref{cst:decode_length_def_1}). Let $d_{i,j,k}\in \{0,1\}$ be the assignment of decode stage $k$ to request $i$ on client $j$ for decode phase, for $i\in \mathcal{I}$, $j\in \mathcal{J}$, and $k\in \mathcal{K}^\text{d}$. Eqs. (\ref{cst:immediate_prefill_decode})-(\ref{cst:prefill-decode-order}) together sets the rule that the decoding phase of a request must immediately follow the consecutive prefilling phase or its previous decoding phase. Eqs. (\ref{cst:d_def_1})-(\ref{cst:loc_def}) relate the assignment decisions among requests, clients, and bins. Eqs. (\ref{cst:var_def_start})-(\ref{cst:var_def_end}) define the domain for each set of variables, respectively.

Eqs (\ref{obj:makespan})-(\ref{cst:var_def_end}) will hereafter be referred to as ``the original model''. We attempted to solve the original model using commercial MIP solvers, such as Gurobi. However, we found it nearly impossible to solve such a large-scale problem directly. The number of requests, denoted by $|\mathcal{I}|$, is around 1,000 in small cases, while in larger cases, it can be up to 10,000. The number of batch sizes or client numbers, denoted by $|\mathcal{J}|$, can reach up to 200. The number of exclusive bins, denoted by $|\mathcal{K}|$, often is on the same order of magnitude as $|\mathcal{I}|$. 
To probe the solving cost, we solved a down-scaled toy case model with merely 100 requests and 20 clients, which took Gurobi more than 3,600 seconds to yield a near-optimal solution without fully closing the dual gap. Solving this problem in its original form is hence evidently impractical and cannot be solved in hours. Therefore, it is necessary to decompose this problem into stages and address it sequentially.

\section{Solution Method}
\label{solution_method}

\begin{figure*}
    \centering
    \includegraphics[width=0.8\linewidth]{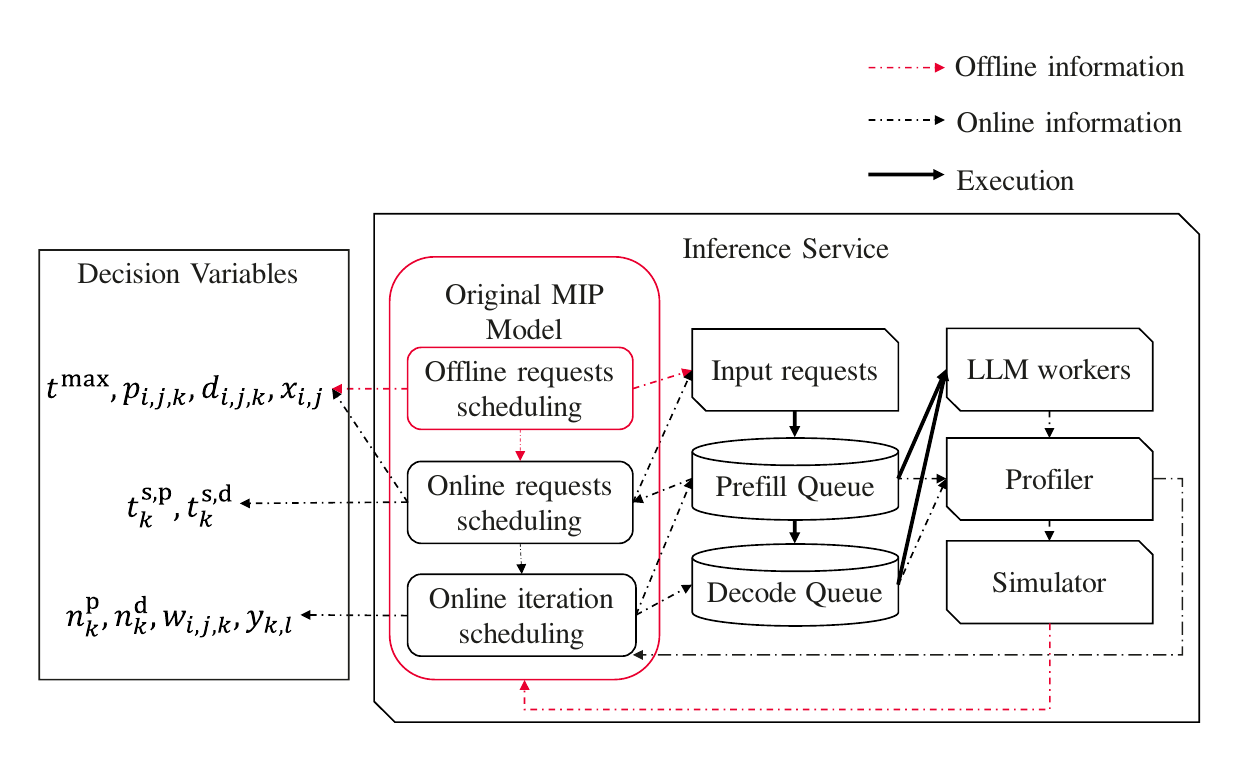}
    \caption{Illustration of Solution Method}
    \label{fig:solution-method}
\end{figure*}

\subsection{Method overview}
The original model, provided by Eqs. (\ref{obj:makespan})-(\ref{cst:var_def_end}), demands a solution method capable of making decisions within 10 milliseconds in an asynchronous cycle, as the decoding time per client batch can be around 50 milliseconds. However, the original model is a large-scale MIP model with over 100,000 integer decision variables and constraints, making it difficult to solve even within several hours. Hence, it is vital to develop an efficient solution method that provides the best possible outcomes within the required time frame. As illustrated in Fig. \ref{fig:solution-method}, we propose a \textit{hybrid offline-online method} that structures the scheduling process for large model inference into two main methods: offline requests assignment and online scheduling. Each  method involves a subset of decision variables of the original model and provides timely solutions at each stage. In the figure, we illustrate how the offline-online information, as well as the decision making given by the scheduling models, is obtained and shared in the system.

\textbf{Offline Requests Scheduling}: In this method, a predetermined batch size of clients determines the number of parallel requests. Each client is allocated a balanced number of requests, resulting in an equitable task distribution. This method considers the assignment decisions in the original model as described in constraints (\ref{cst:makespan_def}), (\ref{cst:d_def_1}), and (\ref{cst:w_def_2})--(\ref{cst:p_def_1}). We isolate this part of the model to demonstrate offline training scenarios, such as RLHF (Reinforcement Learning with Human Feedback) training. In this task, requests are typically given and known in advance. Users can manually send their request prompts to the LLMs and wait to receive the outputs. These tasks can implement offline request assignment methods to achieve better throughput. However, for most scenarios such us user using GPT, using the offline method is still limited since solving the MIP model usually takes 10 minutes or more, which cannot meet the rapid iteration requirements of the LLM online decision-making process. Therefore, it is necessary to develop an online method to fill this gap.

\textbf{Online Scheduling}: The online scheduling process comprises two major parts corresponding to two types of online decisions. The first part, \textit{online requests scheduling}, determines which requests are scheduled in the upcoming bin and identifies the next client to serve once a previous request is completed. The second part, \textit{online iteration scheduling}, decides when to conclude the current decoding bin and start a preemptive prefilling bin to enhance overall utilization rates.

In the online requests scheduling part, heuristic methods are employed to determine the optimal order for processing requests in real-time. This approach considers factors such as task priority, current system load, and anticipated resource availability. By effectively prioritizing requests, the system can minimize waiting times and maximize throughput under dynamic operational conditions. This method emphasizes the implementation of the relaxed solutions provided by job assignment and illustrates the constraints (\ref{cst:bin_def_2})--(\ref{cst:prefill-decode-order}) in the original model.

The online iteration scheduling part aims to minimize idle time on machines by strategically allocating computational resources. By dynamically adjusting prefilling and decoding task priorities based on real-time feedback and system constraints, this method enhances overall system efficiency and responsiveness. This proactive scheduling approach minimizes machine idle time and optimizes the utilization of processing resources, thereby improving the overall performance of large language model inference tasks. This method underscores iteration-based optimization and considers the constraints (\ref{cst:w_def_1})--(\ref{cst:w_def_2}) and (\ref{cst:p_def_2})--(\ref{cst:loc_def}) in the original model.

The overarching goal of this structured approach is to minimize the total inference time for a specific benchmark, thereby maximizing throughput. By integrating thorough data analysis, efficient task allocation, and adaptive online scheduling strategies, this scheduling solution optimizes the performance of LLM inference processes. This holistic approach not only enhances system efficiency but also supports scalability and reliability in handling complex computational tasks.

\subsection{Offline requests scheduling \& theoretical lower bound}
As previously introduced, we begin by examining the offline request assignment decisions within the original model, with a specific focus on the constraints described in (\ref{cst:makespan_def}), (\ref{cst:d_def_1}), and (\ref{cst:w_def_2}--\ref{cst:p_def_1}). This part of the model is isolated to demonstrate offline training scenarios, such as RLHF training. In this scenario, we tackle the Minimizing Makespan Bin Packing Problem to efficiently address the workload balancing challenge. We assume that the output length is predetermined, and that prefill decode stages do not conflict during problem-solving. Nevertheless, in practical applications and simulations used to evaluate performance, we adhere to these constraints by allocating workload to clients without affecting the uncertainty of output length.

In the offline model outlined in Eqs. (\ref{offline_bin_packing_start})--(\ref{offline_bin_packing_end}), we introduce a new parameter, denoted by $T_i \in \mathbb{R}^+$ for $i\in \mathcal{I}$, representing the estimated decode completion time for request $i$. We also introduce a new decision variable, denoted by $t_j \in \mathbb{R}^+$ for $j\in \mathcal{J}$, to indicate the total decoding time for client $j$.

\begin{align}
    \label{offline_bin_packing_start}
    & \min \max_{j\in \mathcal{J}} t_j \\
    & \nonumber \text{s.t.} \\
    & \sum_{j \in \mathcal{J}} x_{i,j} = 1, \text{ for } i\in \mathcal{I}, \\
    & \sum_{i\in \mathcal{I}} x_{i,j} T_i \leq t_j, \text{ for } j\in \mathcal{J}, \\    
    & x_{i,j} \in \{0,1\}, \text{ for } i\in \mathcal{I},\text{ and }j\in \mathcal{J},\\
    \label{offline_bin_packing_end}
    & t^\text{max}, t_j \in \mathbb{R}^+, \text{ for } j \in  \mathcal{J}.
\end{align}

The offline model also provides a method to calculate the theoretical lower bound for a given set of requests, $\mathcal{I}$. In this method, we assume that prefill and decode phases for all the requests can be separated into two groups, and we calculate the optimal inference time for each group.

Let $t^{\text{p}*} \in \mathbb{R}^+$ and $t^{\text{d}*} \in \mathbb{R}^+$ represent the optimal total prefill and total decode times for all the requests in set $\mathcal{I}$, respectively. The value of $t^{\text{d}*}=\max_{j\in \mathcal{J}} t_j$ is obtained from the objective function value from Eqs. (\ref{offline_bin_packing_start})--(\ref{offline_bin_packing_end}). Let $L=\arg\max_{l \in \mathcal{L}} N_l^{\text{cap}}$, and then the largest prefill time across all levels in $T^\text{p}_l$ is denoted by $T^\text{p}_L$. $N^{\text{cap}}_L$ is the number of maximum number of tokens that can be processed in $T^\text{p}_L$. Then, $t^{\text{p}*}$ can be calculated by the following equation. 

\begin{align}
t^{\text{p}*} \geq T^\text{p}_L  \left\lfloor \frac{\sum_{i\in \mathcal{I}} N_i^{\text{p}}}{N^{\text{cap}}_L} \right\rfloor.
\end{align}
It yields a tight theoretical lower bound $T^{\text{LB}}$ as follows.

\begin{align}
\label{theoreticalLB}
T^{\text{LB}} = t^{\text{p}*} + t^{\text{d}*}.
\end{align}

\subsection{Online requests and iteration scheduling}
In this online part of the LLM inference optimization problem, two critical considerations arise. First, we need to determine which request to send to an available client once the previous request is completed. Second, when a round of decoding stage is finished, we must decide whether to send a preemptive prefill stage or continue this decode stage to the LLM workers for the subsequent time frame.

The first issue presents an online scheduling problem, as illustrated by Eqs. (\ref{cst:w_def_1})--(\ref{cst:w_def_2}) and (\ref{cst:p_def_1})--(\ref{cst:p_def_2}). The primary decision in this context is whether to select a new request to override the original assignment in order to achieve better machine utilization.

A sorting and online preemptive method is illustrated in Algorithm \ref{algorithm1}. This online algorithm first selects the future available requests, denoted as $I_j$, for client $j \in \mathcal{J}$. The set $I_j$ is sorted by $N_i^{\text{p}} + N_i^{\text{d}}$, that is, $N_{i_1}^{\text{p}} + N_{i_1}^{\text{d}}>N_{i_2}^{\text{p}} + N_{i_2}^{\text{d}} \forall i_1<i_2 \in I_j$. Then, for each client, the algorithm calculates future requests and counts the expected remaining tokens $remain\_token(j) \text{ for } j\in \mathcal {J}$ to be processed. Idle clients then greedily select the longest request from busy clients to process. This algorithm utilizes offline information on request assignment to provide timely online assignment decisions.

\begin{algorithm}
\caption{Sorting and Online Preemptive Method}
\label{algorithm1}
\begin{algorithmic}
\Require $\{ I_j = \{ i \mid x_{ij} = 1 \}, \forall j \in \mathcal{J} \}$ 
\For{client $j$ in clients $\mathcal{J}$}
    \If{queue for client $j$ is empty and $I_j \neq \emptyset$}
        \State pop $I_j$ to client $j$
        \State $remain\_token(j) \gets remain\_token(j) - (N_i^{\text{p}} + N_i^{\text{d}})$
    \ElsIf{$\max(remain\_token(j)) > 0$}
        \State pop $\arg\max(remain\_token)$ to client $j$
        \State $remain\_token(j) \gets remain\_token(j) - (N_i^{\text{p}} + N_i^{\text{d}})$
    \EndIf
\EndFor
\end{algorithmic}
\end{algorithm}

Continuing from the previous discussion, the second problem involves a sequential decision-making process, as outlined by Eqs. (\ref{cst:makespan_def})--(\ref{cst:prefill-decode-order}). The main challenge here is to deliver timely and efficient decisions in real time. As previously mentioned, each round of decoding takes approximately 50 milliseconds. Thus, it is essential to ensure that decisions are made within 10 milliseconds to sustain system efficiency. To achieve this, we employ the following method to integrate quick decision-making into the process.

This aspect of decision-making corresponds to the following problem.

\begin{align}
    &\nonumber \min t^\text{max} \\
    & \nonumber \text{s.t.} \\
    \label{tmax}
    & t^\text{max} \geq t_{k}^\text{s,d} + n_{k}^\text{d}, \text{ for } k \in \mathcal{K}^\text{d}, \\
    & t^{\text{s}}_{p,k} - (t^{\text{s}}_{d,k-1} + n^{\text{d}}_{k-1}) \geq 0,  \text{ for } k = 2,...,K,\\
    \label{tsdk}
    & t^{\text{s,d}}_{k} - (t^{\text{s,p}}_{k} + n^{\text{p}}_{k}) \geq 0,  \text{ for } k \in \mathcal{K}^\text{p},\\ 
    \label{npk}
    & n^{\text{p}}_{k} \geq \sum_{l\in \mathcal{L}} T^{\text{p}}_l y_{k,l}, \text{ for } k \in \mathcal{K}^\text{p},\\
    & \sum_{i\in \mathcal{I},j\in \mathcal{J}} N_i^{\text{p}} p_{i,j,k} \leq \sum_{l\in \mathcal{L}} N^{\text{cap}}_l y_{k,l}, \text{ for } k \in \mathcal{K}^\text{p}, \\ 
    & \sum_{l\in \mathcal{L}} y_{k,l} = 1, \text{ for } k \in \mathcal{K}^\text{p}, \\
    \label{ndk}
    & n^{\text{d}}_{k} \geq T^{\text{d}} \sum_{i} N_i^{\text{d}} w_{i,j,k}, \text{ for } j\in \mathcal{J}, \text{ and } k \in \mathcal{K}^\text{d},\\
    & d_{i,j,k} - p_{i,j,k} \geq 0, \text{ for } i\in \mathcal{I}, j\in \mathcal{J}, \text{ and } k \in \mathcal{K}.
\end{align}
By combining Eqs. (\ref{tmax}) and (\ref{tsdk}), we derive that $t^{\text{max}} \geq \max_{k \in \mathcal{K}} \left(t^\text{{s,p}}_{k} + n^\text{p}_{k} + n^\text{d}_{k}\right) $. In the context of online decision-making, the start time $t^\text{{s,p}}_{k}$ is typically influenced by the completion time of preceding tasks. The primary objective is to minimize the total time cost of prefill and decode stages. Consequently, we establish the following equation by integrating the calculations of $n^\text{p}_{k}$ and $n^\text{d}_{k}$ from Eqs. (\ref{npk}) and (\ref{ndk}).

\begin{align}
    \min t^\text{max} \geq \max_{k \in \mathcal{K}} \left( t^\text{s,p}_{k} + \sum_{l\in \mathcal{L}} T^\text{p}_l y_{k,l} + T^\text{d} \sum_{i \in \mathcal{I}, j\in \mathcal{J}} N_i^\text{d} w_{i,j,k} \right).
\end{align}
In this problem, the time cost is divided into two components. The cost for adding a prefill task at any point is given by

\begin{align}
    \frac{\partial t^\text{max}}{\partial y_{k,l}} = \sum_{l \in \mathcal{L}} T^\text{p}_l,
\end{align}
and the cost for adding a decode task at the decision-making moment is expressed by

\begin{align}
    \frac{\partial t^\text{max}}{\partial w_{i,j,k}} = T^\text{d} \sum_{i\in \mathcal{I}} N_i^\text{d}.
\end{align}
Thus, our heuristic method for deciding whether to dispatch a prefill or decode stage to the LLM worker involves comparing the prefill cost $C_p = \sum_l T^p_l$ with the waited decode time $C_d = T^d \sum_{i\in \mathcal {I}, j\in \mathcal{J}} N_i^d w_{i,j,k}$. If $C_p \geq C_d$, the algorithm advises continuing with a round of the decode task and waiting for additional prefill tasks; otherwise, the algorithm recommends executing a round of the prefill task.

\section{Numerical Experiment}
\label{numerical-experiment}
\subsection{Experiment settings and baseline}
Before the model is solved by our hybrid method, extensive analysis is conducted to evaluate the time taken by the decode and prefill stages on hardware. This analysis provides crucial insights into the computational demands and performance characteristics of each stage. By quantifying these metrics, such as processing times and resource utilization, the data analysis establishes a solid foundation of empirical data. The data serve as reliable support for subsequent decision-making in optimizing scheduling strategies. The basic experiment setting is given in TABLE \ref{Experiment Settings}.

\begin{table}
    \centering
    \caption{Experiment Settings}
    \begin{tabular}{ccc}
    \toprule
    Parameter     & Number & Brief Description\\
    \midrule
    $|\mathcal{I}|$     & 1319 & The GSM8K dataset \\
    $|\mathcal{J}|$     & 200 & Due to hardware memory limit\\
    $\mathbf{E}(N_i^\text{p})$     & 68.43 & The GSM8K dataset input\\
    $\mathbf{E}(N_i^\text{d})$     & 344.83 & The Llama 65B output\\
    $T^\text{p}$     & 0.13 ms/token & The hardware performance on prefilling\\
    $T^\text{d}$     & 0.21 ms/token & The hardware performance on decoding\\
    \bottomrule

    \end{tabular}
    \label{Experiment Settings}
\end{table}

We demonstrate our improvements by utilizing the GSM8K dataset \cite{cobbe2021training}, a comprehensive collection of mathematical word problems specifically designed to assess the problem-solving and reasoning capabilities of language models. This dataset serves as a benchmark for evaluating both arithmetic and logical reasoning skills, essential attributes for advanced language models. Each problem in the dataset is intricately constructed to simulate real-world situations involving numerical relationships, necessitating that models comprehend the problem contextually and perform accurate calculations to derive the correct solution.

The GSM8K dataset comprises 1,319 unique problems as input requests, with an average input length of 68.43 tokens and a standard deviation of 25.04 tokens. For our experiments, we selected the LLaMA-65B language model due to its open-source nature and wide accessibility, making it a suitable candidate for academic and reproducible research. In our tests, the LLaMA-65B model generated responses averaging 344.83 tokens, with a standard deviation of 187.99 tokens. To ensure consistency and focus on quality responses, we constrained the maximum output length to 512 tokens during testing.

Our computational setup is characterized by a robust hardware configuration, consisting of eight Ascend processing units, each equipped with a maximum memory capacity of 64 GB. This formidable hardware infrastructure is essential for facilitating the efficient processing and testing necessary for our experiments. Additionally, we have assessed the KV cache usage for input in this experiment, establishing baseline settings that are also utilized in practical applications. The current hardware, along with the LLM employed, imposes a memory constraint of 1024 blocks of KV cache. Each block can accommodate a maximum of 128 tokens. For the GSM-8k benchmark, the combined maximum input and output for each request requires five blocks. Consequently, this configuration limits us to a maximum of approximately 200 clients running concurrently, calculated by the expression $1024/5 \approx 200 $.

In our experimental setup, we conduct an estimation of the operation time required for prefill and decode stages using over 400 data groups. We find that both prefill and decode times exhibit a linear relationship with the number of tokens involved. Specifically, the prefill time can be calculated as 0.13 milliseconds per token, plus a fixed overhead of 25 milliseconds. For the decode process, the time required for each batch of clients can be estimated as 0.21 milliseconds per token, with an additional fixed overhead of 29 milliseconds. For instance, when processing a parallel batched decode stage involving 200 clients, where each client produces one token per round, the operation would take approximately $200 \times 0.21 + 29 = 71$ milliseconds. In the case of prefill stages, if a batch consists of inputs totaling 5,000 tokens, the estimated time required would be $5000 \times 0.13 + 25 = 675$ milliseconds.

\begin{figure}
    \centering
    \includegraphics[width=0.99\linewidth]{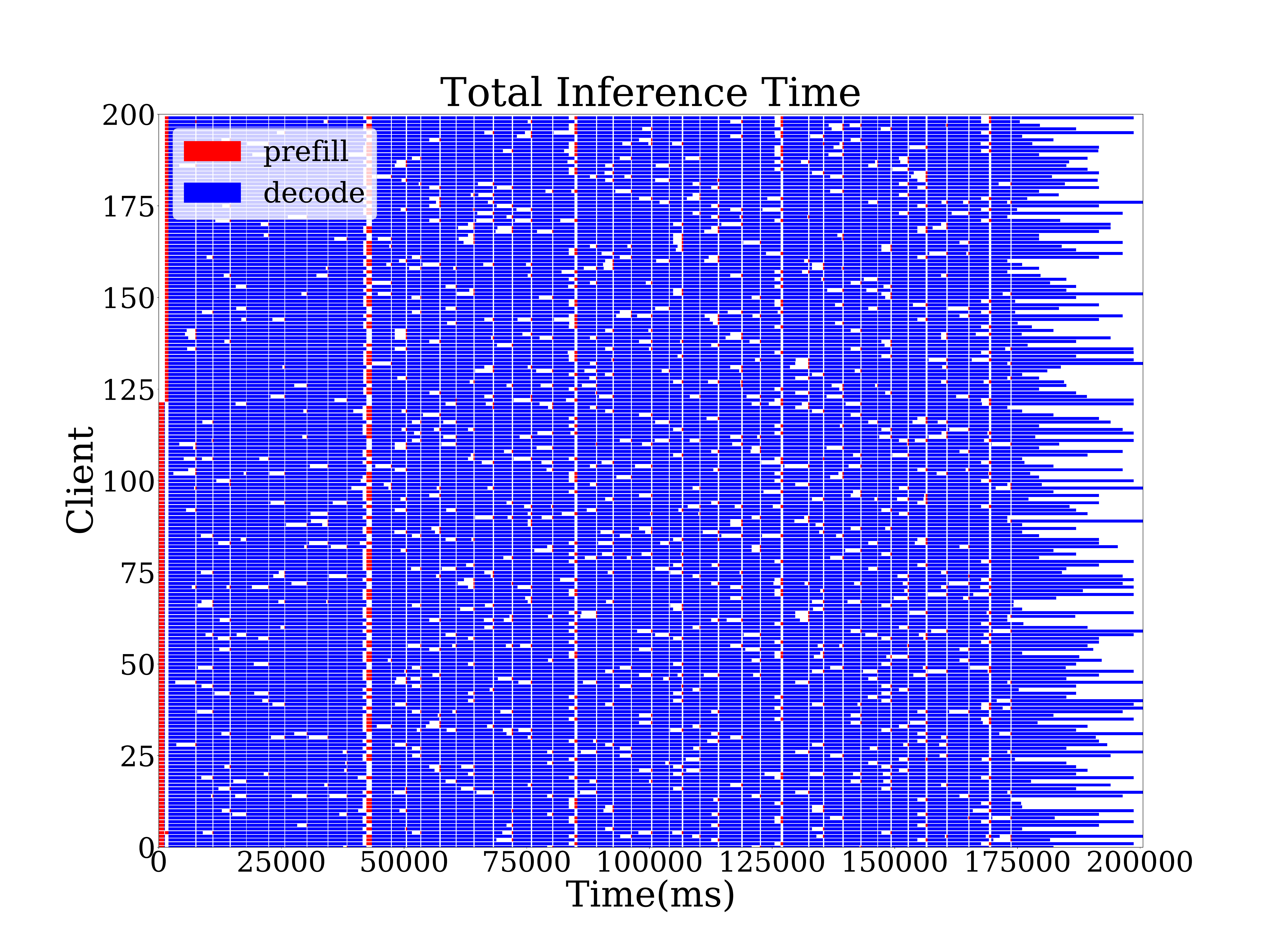}
    \begin{flushleft}
    \vspace{-2em}
    \caption*{Utilization rate: 80.2\%. Total inference time: 201.00 seconds.}
    \end{flushleft}
    \vspace{-2em}
    \caption{Result Gantt: Baseline}
    \label{fig:baseline}
\end{figure}

We present a Gantt chart in Fig. \ref{fig:baseline}, generated from an experiment using real-world data and open source LLM, to illustrate the current state of online inference services without the implementation of our proposed method. This chart demonstrates that, in practical scenarios, a significant number of idle periods, or ``bubbles'', occur when no scheduling strategy is employed. Furthermore, in offline scenarios, if the workload among clients is not evenly distributed, substantial machine idle time is observed after the early completion of some client's tasks. Our analysis of this Gantt chart reveals that the overall machine utilization rate is only 80.2\%.

\subsection{Offline request scheduling result}
\begin{figure}
    \centering
    \includegraphics[width=0.99\linewidth]{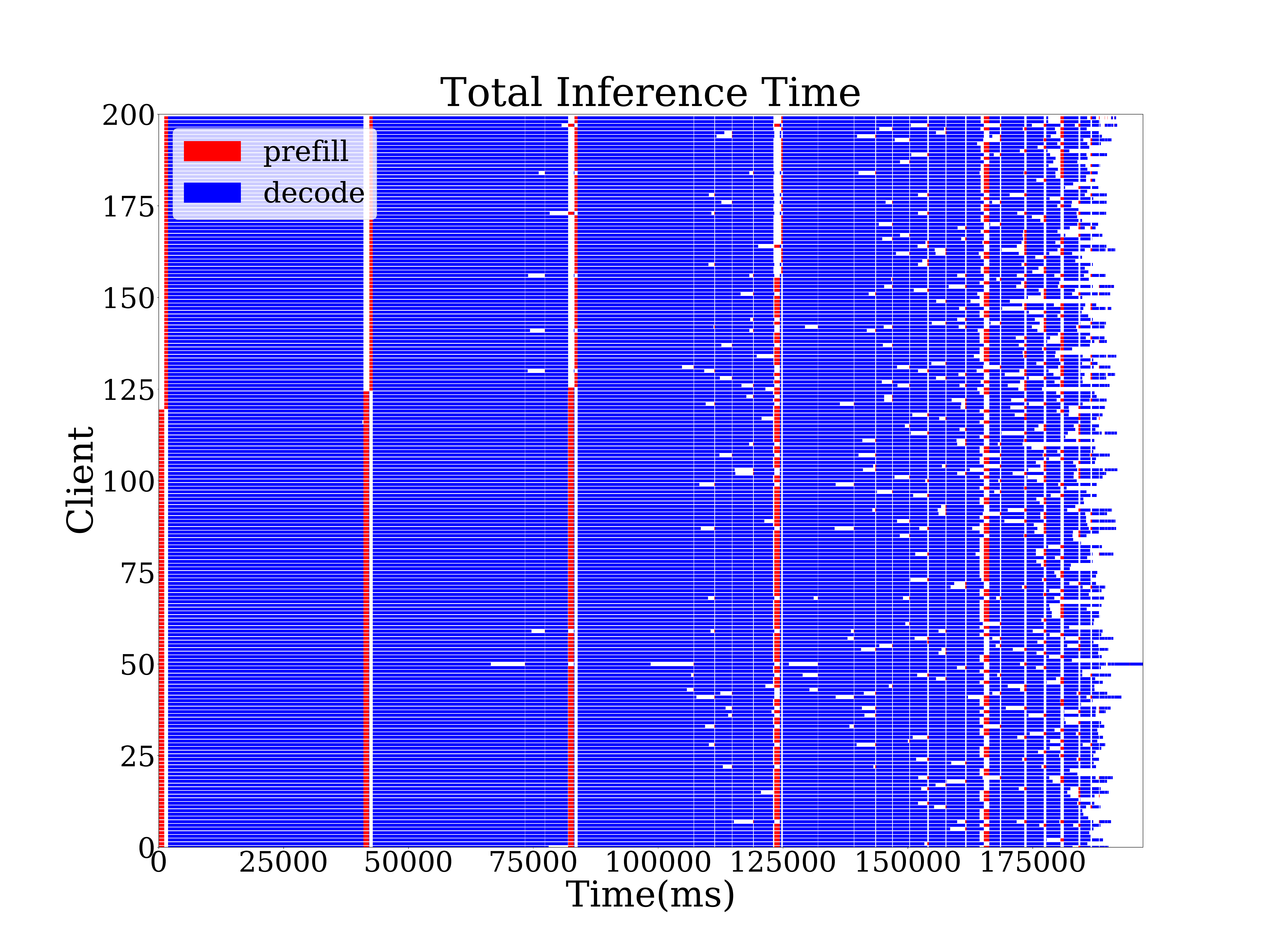}
    \begin{flushleft}
    \vspace{-2em}
    \caption*{Utilization rate: 85.5\%. Total inference time: 197.08 seconds.}
    \end{flushleft}
    \vspace{-2em}
    \caption{Result Gantt: Offline Request Scheduling}
    \label{fig:offline_assignment}
\end{figure}

This offline request scheduling model given by Eqs.  (\ref{offline_bin_packing_start})--(\ref{offline_bin_packing_end}) can be solved using open-source solver SCIP. Due to significantly reduced complexity, optimal solutions can be achieved within 20 minutes comparing to original problem which is not possible to be solved within hours. Although this offline model only addresses workload balancing using estimations of output length, its performance surpasses that of the original version. As illustrated in Fig. \ref{fig:offline_assignment}, the system shows a significant reduction of idle times, and machine utilization is enhanced to 85.5\%. Comparing to the baseline method, this method provides a more balanced request assignment across clients and reduce ``bubbles''. The total inference time can be reduced from 201.00 seconds to 197.08 seconds. Since solving the model still takes relatively long time, we list this method as optional and suggest practitioners use the offline model in typical scenarios such as RLHF training. 

\subsection{Online scheduling results}

\begin{figure}
    \centering
    \includegraphics[width=0.99\linewidth]{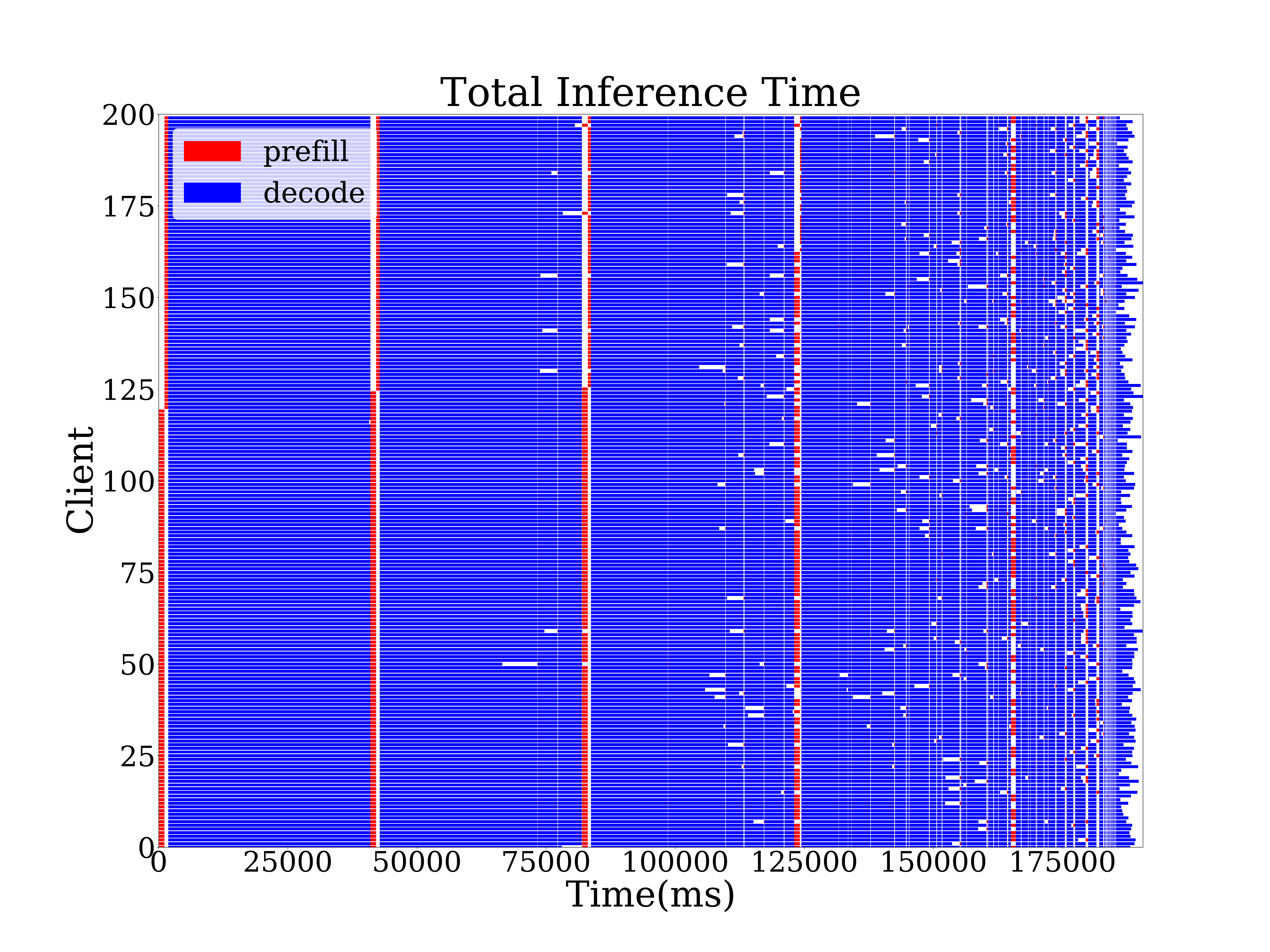}
    \begin{flushleft}
    \vspace{-2em}
    \caption*{Utilization rate: 89.06\%. Total inference time: 190.58 seconds.}
    \end{flushleft}
    \vspace{-2em}
    \caption{Result Gantt: Offline+Online Scheduling}
    \label{fig:online+offline}
\end{figure}

Incorporating online requests and iteration scheduling methods, as depicted in Fig. \ref{fig:online+offline}, results in a marked improvement in total inference time, showing reductions 190.58s compared to 201.00s in the baseline scenario. Additionally, machine utilization is enhanced to 89.06\%. Comparing to offline scheduling method, these two online methods do not require additional computing and can be used for current online inference.

\begin{figure}
    \centering
    \includegraphics[width=0.99\linewidth]{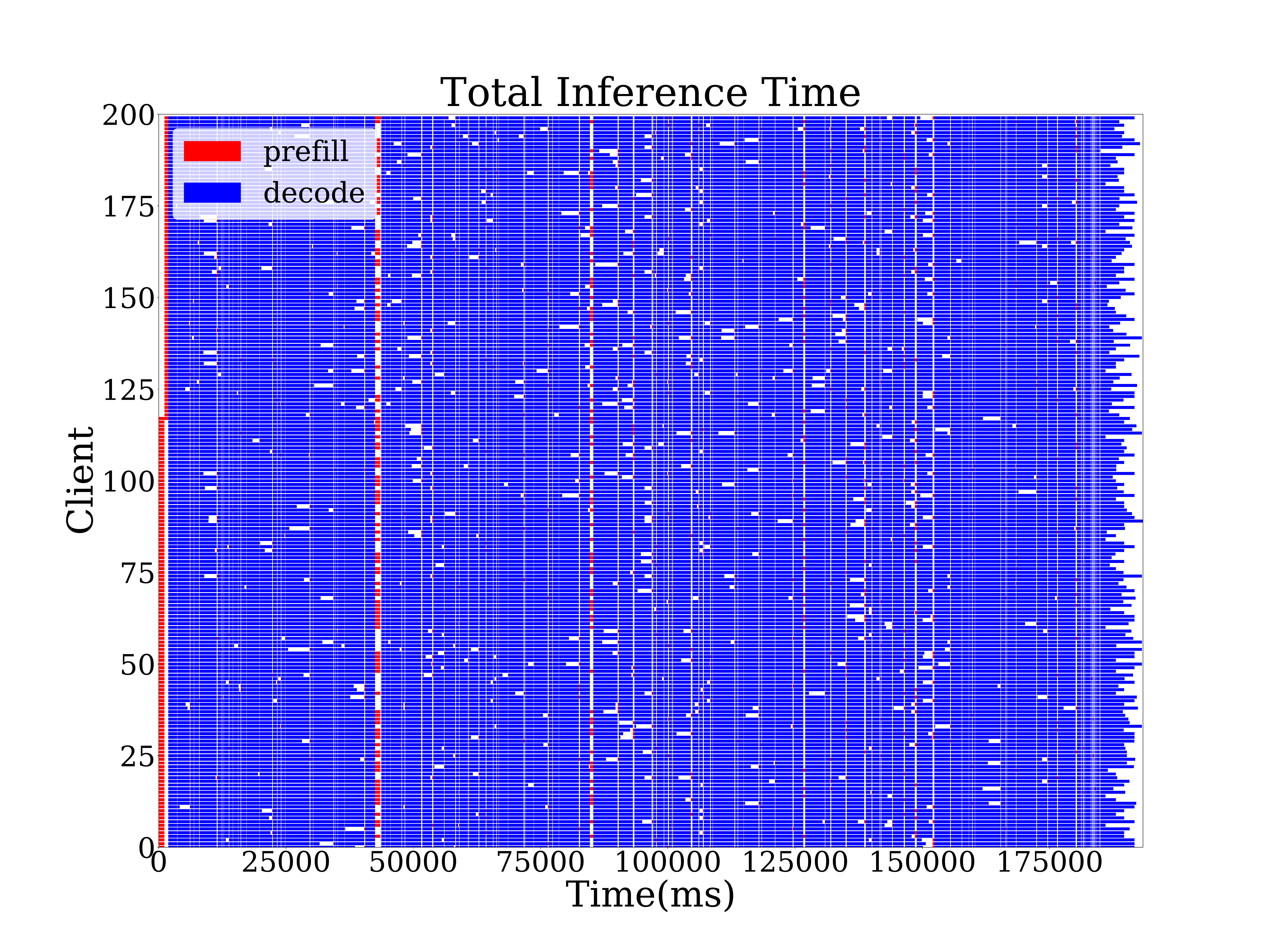}
    \begin{flushleft}
    \vspace{-2em}
    \caption*{Utilization rate: 86.19\%. Total inference time: 193.33 seconds.}
    \end{flushleft}
    \vspace{-2em}
    \caption{Result Gantt: Online only Scheduling}
    \label{fig:online_only}
\end{figure}

In Fig. \ref{fig:online_only}, we present the results obtained using only the online scheduling method, without employing the offline scheduling method. As shown, compared to the baseline, the utilization rate improves to 86.19\%, and the total inference time decreases to 193.33 seconds. These results demonstrate that the online method performs well even in the absence of prior knowledge about requests. This scenario is common in the area of LLM inference.

We also calculate the theoretical lower bound using Eq. (\ref{theoreticalLB}). In the specified numerical case utilizing GSM8K, the theoretical bound is 180 seconds, in which $T^{\text{p}*}=13$ seconds and $T^{\text{d}*}=167$ seconds. In this scenario, we reduce the total inference time from 201.00 seconds in the baseline to 190.08 seconds with the hybrid online-offline method. The gap to the optimal value is thus reduced from \(201 - 180 = 21\) seconds to \(190 - 180 = 10\) seconds, representing a reduction of 52.4\% in this ``primal dual'' gap.


\begin{figure}
    \centering
    \includegraphics[width=0.99\linewidth]{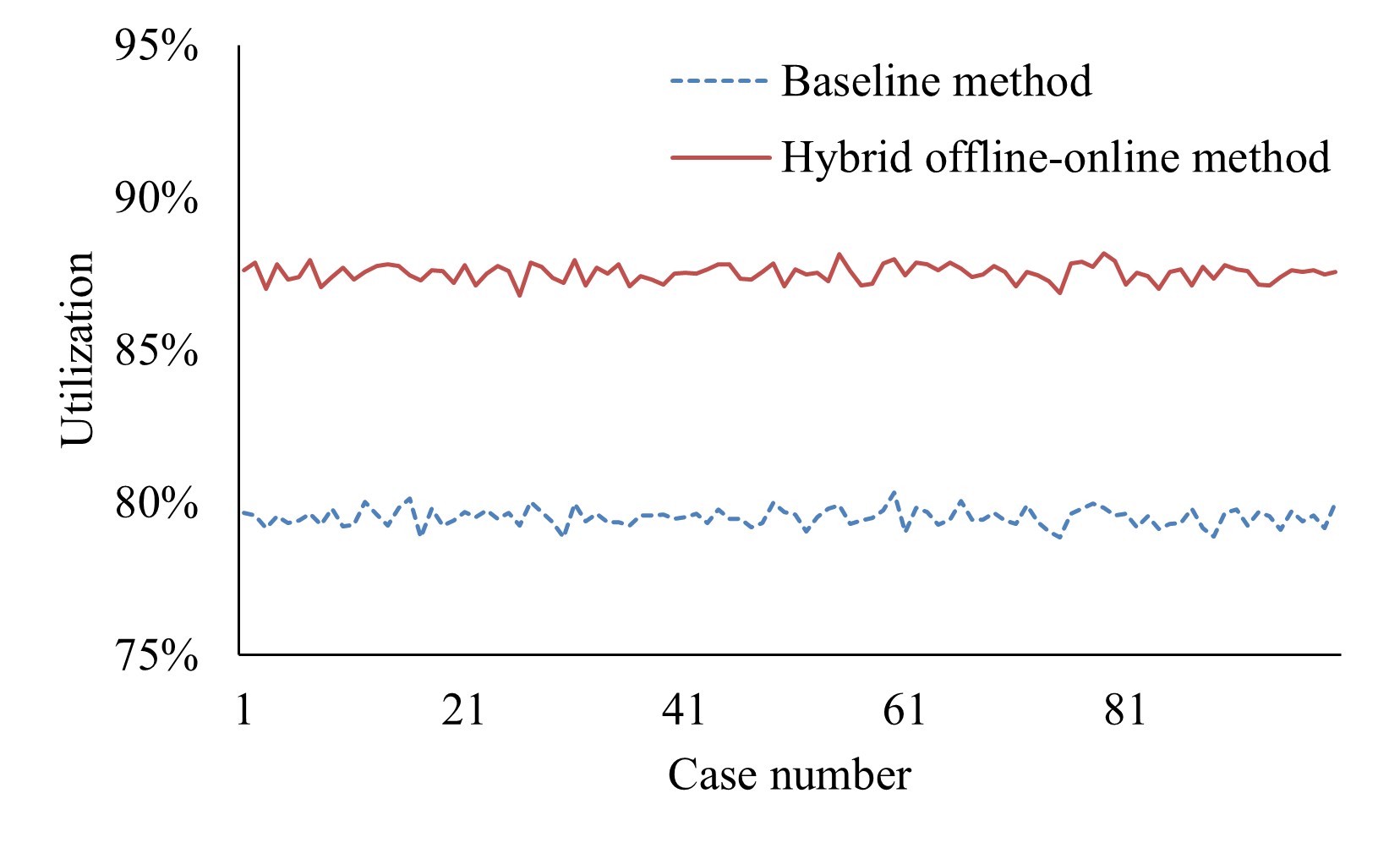}
    \caption{Utilization rate with 100 cases}
    \label{fig:result_numerical_utilization}
\end{figure}

\begin{figure}
    \centering
    \includegraphics[width=0.99\linewidth]{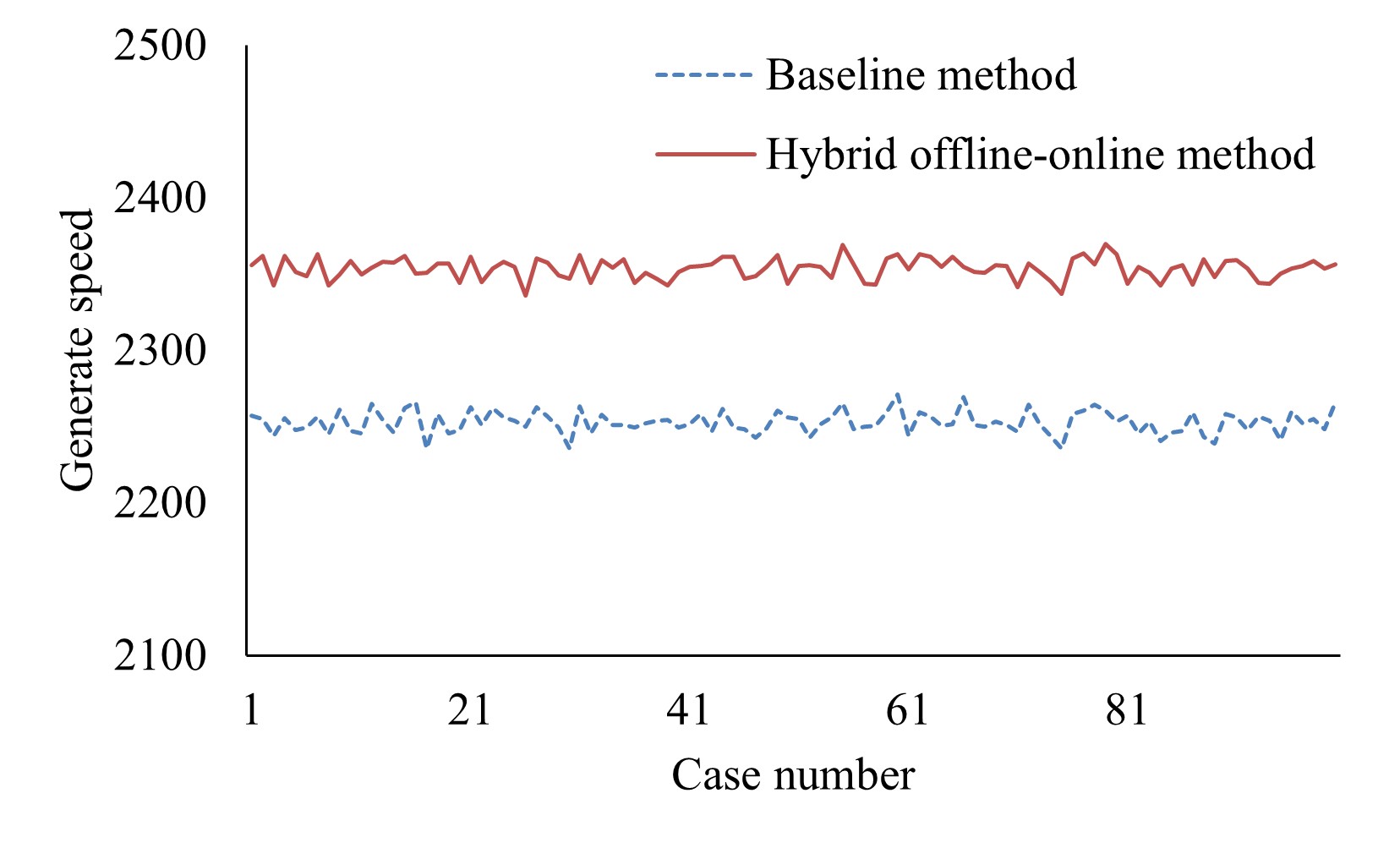}
    \caption{Generate speed with 100 cases}
    \label{fig:result_numerical_generate_speed}
\end{figure}

To better demonstrate the performance of our online scheduling methods, we present a numerical experiment involving 100 cases in Figs. \ref{fig:result_numerical_utilization} and  \ref{fig:result_numerical_generate_speed}. These cases are randomly generated with the input and output length distributions shown in TABLE \ref{Experiment Settings}. As illustrated in the figures, despite some variations across the 100 cases, our hybrid offline-online method consistently outperforms in both utilization and generation speed. The unit for generation speed is tokens per second, indicating how many tokens the LLM can generate each second. On average, our method achieves an 8.0\% improvement in utilization and an increase of 100.63 tokens per second in generation speed.

\section{Conclusion and Future Work}
\label{conclusion}
In this paper, we study the inference optimization problem in the service system when deploying LLMs. To enhance the system throughput and better utilize the hardware, we formulate an MIP model to describe this inference optimization problem. To the best of our knowledge, this is the first formulation of the problem from a scheduling perspective. To tackle the complex and high real-time demands of LLM inference, we introduce a hybrid offline-online method.

In the offline method, we demonstrate how large-scale inference systems can be improved using a Minimizing Makespan Bin Packing Problem and how a theoretical lower bound can be provided. In the online request scheduling and iteration scheduling methods, the solution time efficiency is crucial. We propose a sorting and online preemptive method to more effectively utilize clients that finish early. Then, we focus on the iteration scheduling component of the original model and employ a Lagrangian method to compare the costs of adding a prefill stage versus a decode stage at each iteration. We provide a time efficient heuristic method to determine when to insert a prefill task and interrupt ongoing decoding tasks. 

In real-world experiments, we deploy the LlaMA-65B LLM model and infer the GSM 8K dataset, which includes 1,319 unique math problems. Our offline model increases machine utilization rates from a baseline of 80.2\% to 85.5\%, and reduces the total inference time from 201 seconds to 197 seconds. Utilizing the online scheduling methods, the system utilization rate can be further increased to 89.1\%, and the total inference time for the dataset can be reduced to 191 seconds. As demonstrated, if all our methods are implemented, system throughput can be improved by 5.46\%, and hardware utilization can increase by 11.0\%. A 100-cases study shows that our method consistently outperforms the baseline method and improves the utilization rate by 8.0\% on average.

The future directions of this research can be extended along three key aspects:

\begin{itemize}
    \item Stochastic Model \& Efficient Solution Method: The original MIP model we proposed is a deterministic equivalence formulation. While solving this model is already computationally challenging, developing a stochastic programming model could further enhance its accuracy by better accounting for uncertainties. Additionally, more efficient solution methods for tackling the original MIP model are needed to meet the millisecond-level real-time requirements of the decision-making process.
    \item Reinforcement Learning on Iteration Scheduling: the current iteration scheduling approach relies on a heuristic online method. Notably, the decision-making process in this method involves choosing between two options: prefill or decode. Since online state variables such as prefill task waiting time, the number of decoding clients, expected decoding time, and expected prefill time are relatively easy to derive, a simple reinforcement learning (RL) model could be trained to assist the scheduler in making decisions dynamically. 
    \item Online Hardware Utilization Method: We observed during hardware experiments that the system's hardware is often underutilized when a static number of clients is employed, due to stochastic variations in the output length. In scenarios where dynamic and continuous batching methods are applicable, investigating online decision-making for hardware utilization could further optimize performance. Specifically, determining the optimal number of clients that can be allocated concurrently to the system at any given time could help enhance resource utilization and overall efficiency.
\end{itemize}



%





\ifCLASSOPTIONcaptionsoff
  \newpage
\fi



%



\bibliographystyle{IEEEtran}
\bibliography{main}

%

\begin{IEEEbiography}[{\includegraphics[width=1in,height=1.25in,clip,keepaspectratio]{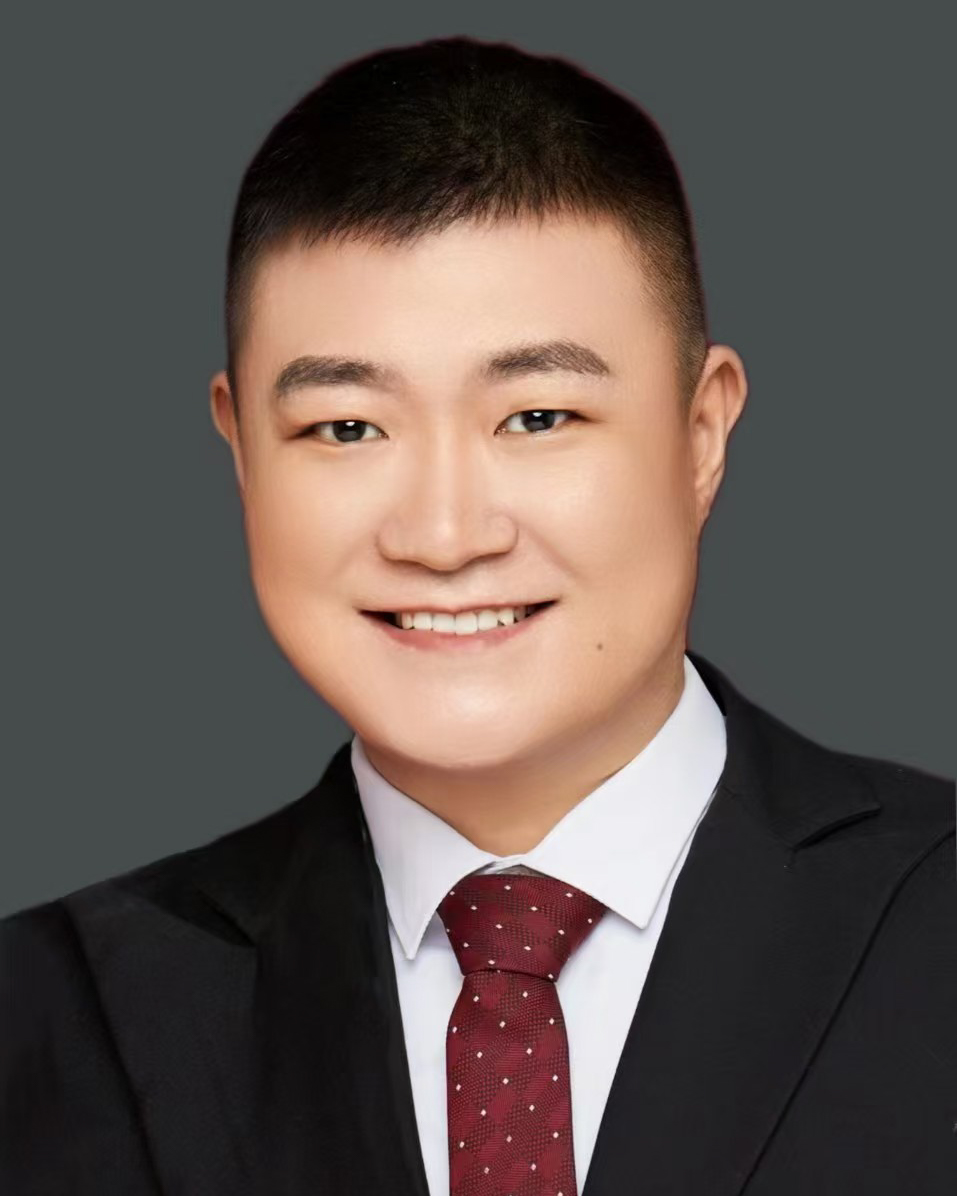}}]{Bowen Pang}
received his Bachelor’s degree and Ph.D. degree from the Department of Industrial Engineering at Tsinghua University, Beijing, China, in 2016 and 2022, respectively. He is currently a researcher at Noah’s Ark Lab, Huawei Technology. His research interests include modeling, analyzing, and solving problems in engineering systems, with applications in healthcare, supply chain, manufacturing, and artificial intelligence. He has been a member of IEEE, IISE, and INFORMS. Please feel free to contact his email pzkaixin@foxmail.com if you are interested in LLM inference optimization area.
\end{IEEEbiography}

\begin{IEEEbiography}[{\includegraphics[width=1in,height=1.25in,clip,keepaspectratio]{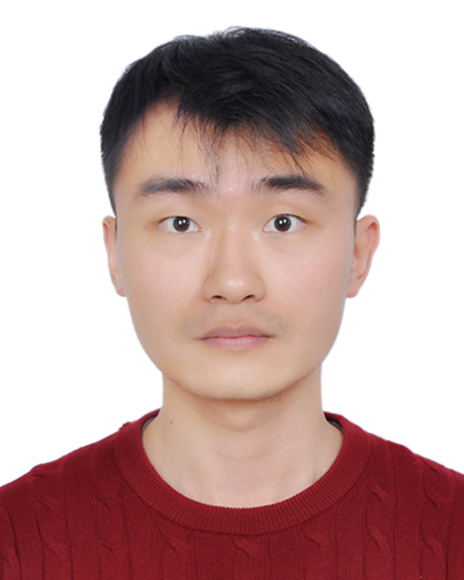}}]{Kai Li}
received the bachelor's degree from YingCai Honors College, University of Electronic Science and Technology of China, Chengdu, China, in 2015, and the Ph.D. degree from the Department of Computer Science and Engineering, Shanghai Jiao Tong University, Shanghai, China, in 2021.
He is currently a researcher at Noah’s Ark Lab, Huawei Technology. 
His research interests include game theory, reinforcement learning, and large language models.
\end{IEEEbiography}

\begin{IEEEbiography}[{\includegraphics[width=1in,height=1.25in,clip,keepaspectratio]{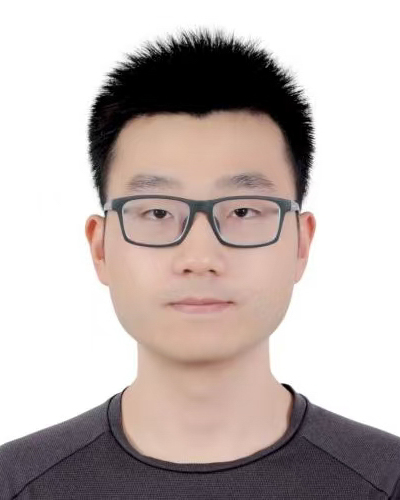}}]{Ruifeng She}
 received his B.S. and Ph.D. degrees from the Department of Civil Engineering of the University of Illinois at Urbana-Champaign, USA, in 2018 and 2023, respectively. He is currently a researcher at Noah’s Ark Lab, Huawei Technology. His research interests include optimization of complex systems, reinforcement learning, and the integration of operations research and machine learning. 
\end{IEEEbiography}

\begin{IEEEbiography}[{\includegraphics[width=1in,height=1.25in,clip,keepaspectratio]{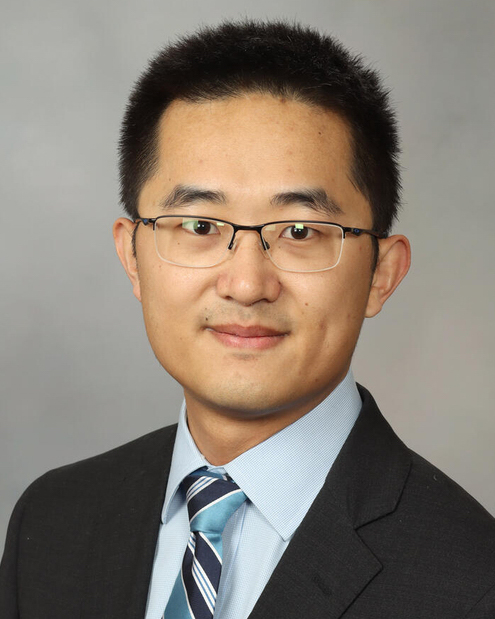}}]{Feifan Wang}
received the bachelor’s degree from the Department of Industrial Engineering, Zhejiang University of Technology, Hangzhou, China, in 2013, the master’s degree from the Department of Industrial and Systems Engineering, Zhejiang University, Hangzhou, China, in 2016, and the Ph.D. degree from the School of Computing, Informatics, and Decision Systems Engineering, Arizona State University, Tempe, AZ, USA, in 2021. He is currently an Assistant Professor with the Department of Industrial Engineering at Tsinghua University, Beijing, China. His research focuses on modeling, analysis, optimization, and control of complex systems, with applications in healthcare delivery systems and production systems. He is a member of IEEE, IISE, and INFORMS. He was a recipient of multiple awards, including the Design and Manufacturing Best Paper Award from the IISE Transactions, the Best Student Paper Award from IEEE CASE, and the Dean’s Dissertation Award from ASU. He has twice been a finalist for the Best Paper Award on Healthcare Automation from IEEE CASE.
\end{IEEEbiography}







\end{document}